\documentclass[reprint,pra,notitlepage]{revtex4-1}
\usepackage{newtxtext}
\usepackage{newtxmath}
\newcommand{\ud}{\mathrm{d}}
\usepackage{amsmath}
\usepackage{amssymb}
\usepackage{wasysym}
\usepackage{bbm}
\usepackage{graphicx}
\usepackage{xcolor}
\usepackage[colorlinks=true,linkcolor=blue,citecolor=blue,urlcolor=blue]{hyperref}
\graphicspath{{fig/}}

\allowdisplaybreaks

\renewcommand{\v}[1]{\mathbf{#1}}
\newcommand{\vv}[1]{\boldsymbol{#1}}

\let\Re\relax
\let\Im\relax
\let\div\relax
\DeclareMathOperator{\Re}{Re}
\DeclareMathOperator{\Im}{Im}
\DeclareMathOperator{\var}{var}
\DeclareMathOperator{\cov}{cov}
\DeclareMathOperator{\curl}{curl}
\DeclareMathOperator{\grad}{grad}
\DeclareMathOperator{\div}{div}

\begin{document}

\title{Seeing beyond the light: Vison and photon electrodynamics in quantum spin ice}
\author{Attila Szab\'o}
\author{Claudio Castelnovo}
\affiliation{TCM Group, Cavendish Laboratory, University of Cambridge, Cambridge CB3 0HE, United Kingdom}

\begin{abstract}
Understanding the nature and behaviour of excitations in quantum spin liquids, and in topological phases of matter in general, is of fundamental importance and has proven crucial for experimental detection and characterisation of candidate materials. Current theoretical and numerical techniques, however, have limited capabilities, especially when it comes to studying gapped excitations. 
Here, we propose a semiclassical numerical method to study systems whose spin liquid behaviour is underpinned by perturbative ring-exchange Hamiltonians. Our method can readily access both thermodynamic and spectral properties. 
We focus in particular on quantum spin ice and its photon and vison excitations. After benchmarking the method against existing results on photons, we use it to characterise visons and their thermodynamic behaviour, which remained hitherto largely unexplored. We find that visons, in contrast to spinons in classical spin ice, form a weak electrolyte: vison pairs are the dominant population at low temperatures. This is reflected in the behaviour of thermodynamic quantities, such as pinch point motifs in the relevant correlators. Visons also appear to strongly hybridise with the photon background, a phenomenon that affects the way these quasiparticles may show up in inelastic response measurements. 
Our results demonstrate that the method, and generalisations thereof, can substantially help our understanding of quasiparticles and their interplay in quantum spin ice and other quantum spin liquids, quantum dimer models, and lattice gauge theories in general. 
\end{abstract}

\maketitle
%
%

\section{Introduction}

Quantum spin liquids (QSL) are topological phases of frustrated magnetic materials in which quantum fluctuations prevent magnetic order even at zero temperature~\cite{Savary2017QuantumReview,Balents2010SpinMagnets}. This phenomenon is often accompanied by exotic behaviour, including emergent gauge symmetries and fractionalised quasiparticle excitations. 
Understanding and characterising such systems has attracted substantial interest in recent years. A particularly timely and important problem is identifying experimentally relevant signatures of QSL behaviour, in light of several candidate materials that have been proposed of late, but remain to be confirmed and characterised. 

Many such phases lend themselves to gauge theoretic descriptions that are reflected in their quasiparticle content, which can in turn give rise to salient features in dynamical spin structure factors and other response properties~\cite{Knolle2019ALiquids}. Understanding these features is both of fundamental importance and a promising diagnostic tool for QSL materials. Indeed, studying the dynamical structure factor in the Kitaev honeycomb model~\cite{Kitaev2006AnyonsBeyond}, for example, has recently led to compelling evidence of a possible QSL phase in $\alpha$-RuCl$_3$ in a magnetic field~\cite{Banerjee2016ProximateMagnet,Banerjee2017Neutron-RuCl3,Do2017Majorana-RuCl3,Banerjee2018Excitations-RuCl3,Kasahara2018MajoranaLiquid}. 

In general, however, response and equilibration properties of quantum spin liquids are far from being well understood. Analytical results are a tall order away from exactly solvable models and numerical techniques that access the excitation spectrum of strongly correlated many-body systems in two and three dimensions have limited capability, especially when facing systems with gapped quasiparticles that are well defined only in a coarse-grained sense. 

A fruitful avenue that has been used to study quantum magnets, including some QSLs, proceeds by approximating quantum spins as classical ones (see for instance Refs.~\cite{Moessner1998PropertiesAntiferromagnet,Conlon2009SpinAntiferromagnets,Conlon2010AbsentAntiferromagnet,Taillefumier2014SemiclassicalLattice,Robert2015SpinYb2Ti2O7,Taillefumier2017CompetingExchange,Robert2008PropagationAntiferromagnet,Schnabel2012FictitiousLattice} on pyrochlore and kagome antiferromagnets and Refs.~\cite{Baskaran2007ExactModel,Baskaran2008Spin-SFunctions,Samarakoon2017ComprehensiveLiquid,Samarakoon2018ClassicalModel} on Kitaev honeycomb systems). 
Removing quantum correlations and uncertainties allows performing Monte Carlo sampling and time evolution more efficiently as well as accessing a wider variety of observables than usually possible with fully quantum methods. 
The starting points of these studies are nearest neighbour bilinear exchange Hamiltonians. 
While this is a successful approach for certain spin liquid phases, QSL behaviour is often underpinned by perturbative ring-exchange processes that induce matrix elements between classically degenerate spin configurations (e.g., in quantum spin ice~\cite{Hermele2004PyrochloreMagnet} and resonant valence bond phases~\cite{Rokhsar1988SuperconductivityGas,Moessner2001ResonatingModel}). 
A direct semiclassical representation of the original Hamiltonian in these cases results in classical phases that do not capture the QSL physics -- this renders the approach inapplicable. 
Indeed, retaining spin liquid behaviour in both semiclassical simulations and large-$S$ path integral calculations depends crucially on a connected continuum of low-energy states between which low-temperature dynamics remains possible. 

In this paper, we develop a semiclassical numerical method to investigate quantum spin liquids that hinges upon a large-$S$ description of the effective ring exchange Hamiltonian. Our approach effectively samples the finite-temperature path integral formulation of the problem using classical Monte Carlo and dynamical simulations to obtain both thermodynamic and spectral properties, thus overcoming previous limitations of semiclassical simulations~\cite{Taillefumier2017CompetingExchange,Onoda2011QuantumOxides}. 

We demonstrate the validity and capability of our approach by studying a class of highly anisotropic quantum spin liquids on the pyrochlore lattice called quantum spin ice (QSI)~\cite{Gingras2014QuantumMagnets}. QSI phases are described by a compact $U(1)$ gauge theory analogous to lattice quantum electrodynamics (QED)~\cite{Hermele2004PyrochloreMagnet}. Similarly to ordinary QED, QSI exhibits linearly dispersing gapless photon modes, as well as gapped electric charges (spinons)~\cite{Note11}.
\footnotetext[11]{There is some ambiguity around the names of emergent fields in the literature. Following the lattice gauge theory convention, we call the emergent vector potential magnetic, so that the description matches the terminology of standard QED. However, this makes spinons sources of electric field, which is at odds with them being sources of the physical magnetic field in dipolar spin systems and real materials.}
Furthermore, the compact nature of the theory allows for Dirac quantised~\cite{Dirac1931QuantisedField} gapped magnetic monopoles [{$U(1)$} visons]. 
While no material has conclusively been classified as QSI to date, evidence points to, for instance, several praseodymium-based pyrochlores as promising candidates~\cite{Rau2019FrustratedPyrochlores}. 

This picture of QSI is consistent with quantum Monte Carlo (QMC) studies~\cite{Benton2012SeeingIce,Shannon2012QuantumStudy,Kato2015NumericalCrossover,Huang2018DynamicsIce,Banerjee2008UnusualLattice,Lv2015CoulombLattice}, but the current understanding of its gapped quasiparticles is limited. Very few experimentally relevant signatures of visons have been proposed~\cite{Chen2016MagneticYb2Ti2O7,Chen2017DiracsClassification}, and they remain elusive to QMC studies, which often suffer from the sign problem~\cite{Shannon2012QuantumStudy,Kato2015NumericalCrossover} and have only limited capability to obtain excitation spectra and dynamics~\cite{Huang2018DynamicsIce}. 
Rigorous analytic treatment of spin-$1/2$ systems is a tall order in all but the simplest cases, and the gauge theory picture is based on analytical soft-spin~\cite{Hermele2004PyrochloreMagnet,Benton2012SeeingIce} or large-$S$~\cite{Kwasigroch2017SemiclassicalIce} expansions which in turn struggle to access gapped excitations quantitatively~\cite{Michal}. Mean field and slave boson approaches~\cite{Savary2012CoulombicPyrochlores,Lee2012GenericIce,Savary2013SpinIce,Hao2014BosonicIce} as well as numerical linked cluster calculations~\cite{Benton2018QuantumLiquid} have been successful in capturing potential phases exhibited by QSI Hamiltonians and transitions between them; however, they have not been equally informative about their excitation spectra.
Finally, exact diagonalisation for three-dimensional strongly correlated quantum systems is limited to very small system sizes~\cite{Onoda2011QuantumOxides,Udagawa2019SpectrumIce}. 
New tools to study excitations in QSI and related systems are therefore in high demand. 

Semiclassical techniques have been applied to the archetypal bilinear QSI model~\cite{Taillefumier2017CompetingExchange}; however, these models order in a $\sigma^z$-polarised classical spin ice (CSI) state in the QSI limit where the Ising term is dominant~\cite{Taillefumier2017CompetingExchange,Onoda2011QuantumOxides}. By taking perturbative ring-exchange processes~\cite{Hermele2004PyrochloreMagnet} into account explicitly, our method is able to capture the spin liquid phase of QSI.
As a benchmark, we derive the dispersion of its photonic modes and show that it is in excellent agreement with the prediction of large-$S$ field theory~\cite{Kwasigroch2017SemiclassicalIce}, and with QMC results on the original spin-1/2 system~\cite{Benton2012SeeingIce,Huang2018DynamicsIce}. 

The majority of the paper is devoted to novel results on the $U(1)$ visons that remained elusive in previous studies of quantum spin ice. We consider both the energetics of zero-temperature metastable vison states, as well as thermal ensembles of visons and photonic excitations. We are able to obtain their bare energy cost and long range Coulomb interaction, borne out of the semiclassical equivalent of the quantum kinetic energy in a purely short ranged Hamiltonian. 
Contrary to spinons, both the energy cost and the interaction strength of visons are controlled by the same energy scale and therefore their relative strength is fixed. We find that visons are in the weak electrolyte limit, where their interaction is strong enough to make nearest neighbour pairs energetically favourable over isolated visons. This has important consequences, for instance, if one aims to develop effective models for visons in QSI~\cite{Chen2017DiracsClassification,Chen2016MagneticYb2Ti2O7,Michal}, since in thermodynamic equilibrium, dilute isolated visons only occur in a relatively dense environment of vison pairs. 
We demonstrate that the weak electrolyte behaviour is reflected in thermodynamic properties of the system, such as the blurring of pinch points of magnetic field correlators as a function of temperature. 
Importantly, we also observe a strong interplay between isolated visons, the aforementioned vison pair plasma, and photons, which significantly affects the thermodynamic vison density. This is possibly a semiclassical reflection of quantum hybridisation of photon and vison excitations in QSI, which would have a substantial impact on the possibility to detect visons experimentally in inelastic response measurements. 
More specifically, the dressing of isolated visons by vison pairs suggests an intriguing analogy with particle--antiparticle bubbles in the strong coupling problem in QED -- an aspect that certainly warrants investigating in future work. 

The present work can readily be extended to investigate photon and vison dynamics, e.g., in relation to magnetic noise and transport measurements, as well as to include spinon excitations. Moreover, our approach is not limited to quantum spin ice systems and can be straightforwardly generalised to other QSLs underpinned by perturbative ring exchange processes, including valence bond and quantum dimer models (e.g., following the route proposed in Ref.~\cite{Szabo2019GeneralisedConstraints}). 

The rest of the paper is organised as follows. 
We discuss our method in general terms in Sec.~\ref{section: simulation}. Section~\ref{section: qsi} introduces the QSI model and describes how the method was applied to it. Benchmarking results on photon modes are presented in Sec.~\ref{section: photon}. Metastable vison configurations are studied at low photon densities in Sec.~\ref{section: bare vison}, while Sec.~\ref{section: thermo} deals with thermodynamic properties of the semiclassical QSI model in the presence of both excitations. Conclusions are drawn in Sec.~\ref{section: conclusion}.
%
%

\section{Semiclassical simulation of quantum spin liquids}
\label{section: simulation}

Consider a Hamiltonian that remains in a QSL phase in the limit of large $S$, that is to say, its eigenstates are massively entangled in any local basis~\cite{Savary2017QuantumReview}. 
In a path integral representation, this entanglement corresponds to the interference of a continuum of equivalent trajectories, related to each other by gauge symmetry~\cite{Auerbach1994InteractingMagnetism}. The interference itself is a defining feature of spin-1/2 QSLs, since it accounts for differences between related classical and quantum spin liquids~\cite{Benton2012SeeingIce}. For large $S$, however, quantum fluctuations become unimportant and interference effects are only apparent at the lowest temperatures. Therefore, a large-$S$ QSL is generally indistinguishable from a classical spin liquid (CSL) characterised by a massive degeneracy of the least action trajectories of the quantum path integral. 

It is now straightforward to obtain static correlation functions of the large-$S$ QSL by Monte Carlo sampling the CSL Boltzmann distribution given by $e^{-\beta H[\v n]}$, with the formally identical Hamiltonian $H[\v n]$ understood to act on unit vectors ${\v n} \equiv \{ {\v n}_i \}$ rather than quantum spins. Such sampling can be done more efficiently than quantum Monte Carlo and never suffers from a sign problem; furthermore, it naturally captures gapped excitations which are hard to treat in analytic large-$S$ calculations~\cite{Kwasigroch2017SemiclassicalIce,Michal}.

The time evolution of the QSL is described by a real time path integral with  action
\begin{equation}
    \mathcal{S} = S\sum_i \int_0^T \!\! \ud t\int_0^1 \! \ud s\,  \v n_i\cdot(\partial_t\v n_i)\times (\partial_s\v n_i) 
    - \int_0^T \!\! \ud t\, H[\v n(t)] \, ,
    \label{eq: simulation: path integral}
\end{equation}
where the first term is the Berry phase for real time evolution of spins~\cite{Auerbach1994InteractingMagnetism}, and $\v n_i$ is a unit vector specifying the spin coherent state. 
As discussed above, this path integral can be regarded in the large-$S$ limit as an incoherent superposition  of independent least action trajectories. For each of these, varying the action~\eqref{eq: simulation: path integral} gives 
\begin{equation*}
    S \v n_i\times(\partial_t\v n_i) = \frac{\partial H[\v n]}{\partial\v n_i}
\end{equation*}
or, equivalently,
\begin{equation}
    \partial_t \vv \sigma_i =  \frac{\partial H[\vv \sigma]}{\partial\vv \sigma_i} \times \vv \sigma_i. 
    \label{eq: simulation: least action EoM}
\end{equation}
In the second line, we used the fact that $\v n_i^2=1$ at all times and thus $\v n_i\cdot\partial_t\v n_i=0$~\cite{Auerbach1994InteractingMagnetism}, and we re-expressed the equation in terms of the expectation values of the spin operators $\vv\sigma_i = S \v n_i$, as customary. Equation~\eqref{eq: simulation: least action EoM} can also be derived from Ehrenfest's theorem using $[\sigma^\mu,\sigma^\nu] = i\varepsilon^{\mu\nu\lambda}\sigma^\lambda$; the semiclassical equations follow by replacing operators with their expectation values.

Simulating the large-$S$ dynamics of the system now involves solving the differential equation~\eqref{eq: simulation: least action EoM}; however, with no quasiparticles and infinitesimal zero point fluctuations, that solution would be trivial at $T=0$. Using finite temperature Monte Carlo configurations as initial conditions of time evolution is a natural prescription for finding dynamic correlators of a classical spin liquid; in the large-$S$ path integral language, this essentially samples least-action trajectories of a finite-temperature (e.g., Keldysh) path integral.

Finally, it is important to understand how QSL quasiparticles might appear in semiclassical simulations. These are often due to canonical quantisation -- for example, emergent photons are quanta of lattice electromagnetic modes and integer spinon numbers are set by the quantisation of angular momentum. In the semiclassical limit, such quantisation is irrelevant, but while individual quasiparticles disappear, we anticipate that equivalent classical normal modes survive and their frequency dispersion remains indicative of the original quasiparticle. By contrast, certain quasiparticles [e.g., {$U(1)$} visons in QSI] are due to $2\pi$-ambiguities of phases: Since angles remain quantised even in a semiclassical setting, such particles survive as gapped excitations in our simulations.
%
%

\section{Quantum spin ice}
\label{section: qsi}

\begin{figure}
    \centering
    \includegraphics[width=3.375in]{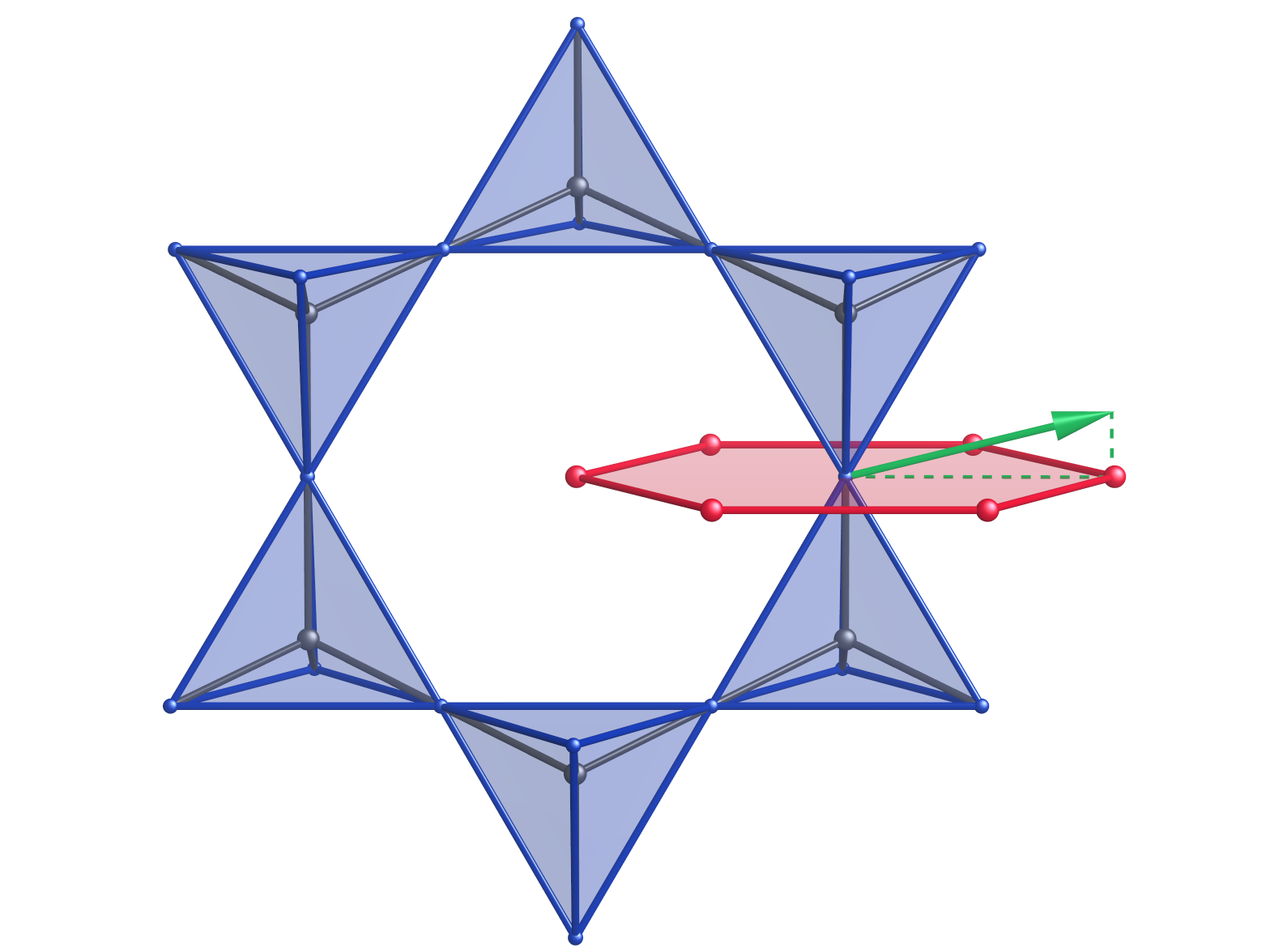}
    \caption{The pyrochlore lattice consists of the bond midpoints of a diamond lattice (black); due to the coordination of the latter, pyrochlore sites form corner-sharing tetrahedra (blue). 
    The shortest closed loops of the pyrochlore lattice are regular hexagonal plaquettes whose centres (red dots) themselves form a dual pyrochlore lattice; the plaquettes of this lattice (red hexagon) are in turn centred on the original sites.
    In spin ice models, Ising (CSI) or Heisenberg (QSI) spins live on each pyrochlore site. In our semiclassical simulation, the anisotropy of the Hamiltonian~\eqref{eq: qsi: ring exchange Hamiltonian} results in mostly easy-plane spins for which $\sigma^z$ (the component along the $\langle 111\rangle$ direction of the corresponding diamond bond) is a small fluctuation.}
    \label{fig:hexagons}
\end{figure}

In the rest of the paper, we focus on quantum spin ice (QSI), a QSL model defined on the pyrochlore lattice (see Fig.~\ref{fig:hexagons}). It is based on the classical spin ice (CSI) model of Anderson~\cite{Anderson1956OrderingFerrites},
\begin{equation}
    H = J \sum_{\langle ij\rangle} \sigma_i^z \sigma_j^z = \frac{J}2 \sum_t \bigg(\sum_{i\in t} \sigma_i^z\bigg)^2 + \mathrm{const.},
    \label{eq: qsi: csi}
\end{equation}
where $t$ runs over the corner-sharing tetrahedra of the pyrochlore lattice. As manifest from the second form,~\eqref{eq: qsi: csi} is minimised if the total $\sigma^z$ around each tetrahedron is zero. This is analogous to Pauling's ``ice rules'' for water ice and gives rise to extensive ground state degeneracy and long-range correlated disorder~\cite{Bramwell2001SpinMaterials}. The elementary excitations of the model are individual tetrahedra violating the ice rules that take the form of fractionalised deconfined spinons~\cite{Castelnovo2012SpinOrder}.

The classical spin ice model~\eqref{eq: qsi: csi} is an Ising model and, as such, its quantum dynamics is trivial. Off-diagonal matrix elements between the classical ground states can be introduced by transverse interactions. Perhaps the simplest such model is the anisotropic XXZ Hamiltonian~\cite{Hermele2004PyrochloreMagnet}
\begin{equation}
    H = \sum_{\langle ij\rangle} J \sigma_i^z\sigma_j^z - \frac{J_\perp}2 (\sigma_i^+\sigma_j^- + \sigma_i^-\sigma_j^+),
    \label{eq: qsi: nn Hamiltonian}
\end{equation}
where $J_\perp\ll J$. The ground state of~\eqref{eq: qsi: nn Hamiltonian} is expected to be a superposition of two-in-two-out spin states, as the Ising term remains dominant; since there is an extensive number of such states, this ground state is massively entangled. 

In the no-spinon sector, the transverse term only acts perturbatively, by flipping closed loops of spins: This can be thought of as creating a virtual pair of spinons and carrying them around the loop. Since the shortest closed loops in the pyrochlore lattice are hexagonal plaquettes, the lowest-order nontrivial perturbative Hamiltonian is
\begin{equation}
    H_\mathrm{eff} = -\frac{g}2 \sum_{\hexagon} \left(\sigma_1^+\sigma_2^-\sigma_3^+\sigma_4^-\sigma_5^+\sigma_6^- + \mathrm{H.c.}\right),
    \label{eq: qsi: ring exchange Hamiltonian}
\end{equation}
where $g = 3J_\perp^3/J^2$~\cite{Hermele2004PyrochloreMagnet}. 

Using soft-spin~\cite{Hermele2004PyrochloreMagnet} or large-$S$~\cite{Kwasigroch2017SemiclassicalIce} approximations, the ring exchange Hamiltonian~\eqref{eq: qsi: ring exchange Hamiltonian} can be mapped onto a compact $U(1)$ lattice gauge theory, where $\sigma^z$ takes the role of the electric field $E$ and the phase of $\sigma^+$ that of the vector potential $A$~\cite{Note11}. 
In particular, the Hamiltonian is invariant under the gauge transformation 
\begin{equation}
    A_{ab}\to A_{ab} + \chi_a-\chi_b,
    \label{eq: qsi: gauge invariance}
\end{equation}
where the $\chi$ are arbitrary angles for each tetrahedron and the subscript $ab$ denotes the spin belonging to neighbouring tetrahedra $a$ and $b$.
This mapping indicates that quantum spin ice hosts gapless photon excitations, while spinons become quantised, gapped sources of the electric field. Furthermore,  the compact gauge theory also supports  $2\pi$-quantised sources of the emergent magnetic field $B=\curl A$, since $B$ itself is only defined modulo $2\pi$. These magnetic charges are called {$U(1)$} visons: Like spinons, they are gapped [by an $O(g)$ gap] and interact through a  Coulomb interaction mediated by photons~\cite{Hermele2004PyrochloreMagnet}. This picture has partly been confirmed, among others, by quantum Monte Carlo studies of the spin-1/2 model~\cite{Benton2012SeeingIce,Shannon2012QuantumStudy,Kato2015NumericalCrossover,Huang2018DynamicsIce,Banerjee2008UnusualLattice,Lv2015CoulombLattice}; however, current analytical and numerical techniques suffer from shortcomings when it comes to investigating its gapped excitations.

It is important to note that the perturbative analysis leading to~\eqref{eq: qsi: ring exchange Hamiltonian} is only valid for $S=1/2$, where diagonal perturbations to the Hamiltonian are equal for all classical ground states. For $S>1/2$, these terms lift the degeneracy of classical ground states and thus disrupt the spin ice phase. As $S\to\infty$, the Ising and transverse terms favour easy-axis and easy-plane arrangements, respectively, leading to energetic competition: In the limit $J_\perp\ll J$, the Ising term wins and the system settles in a classical spin ice state with small transverse spin components~\cite{Taillefumier2017CompetingExchange,Onoda2011QuantumOxides}. These states are stable against transverse fluctuations, removing any dynamics analogous to QSI. The ring exchange Hamiltonian~\eqref{eq: qsi: ring exchange Hamiltonian}, however, remains in a QSL phase for arbitrary $S$~\cite{Hermele2004PyrochloreMagnet,Kwasigroch2017SemiclassicalIce}: Since it only contains transverse terms, its classical ground state is easy-plane and the electric field appears as small fluctuations of $\sigma^z$, which are amenable to continuous time evolution (cf.~Fig.~\ref{fig:hexagons}).

Using~\eqref{eq: simulation: least action EoM}, we developed semiclassical equations of motion for the ring exchange Hamiltonian~\eqref{eq: qsi: ring exchange Hamiltonian}. 
These can be written in the familiar Larmor form $\dot{\vv \sigma}_i = \vv\sigma_i\times\v h_i$ with effective field
\begin{align}
    \v h_i &= (\Re h_i,\Im h_i,0); &
    h_i &= g \sum \sigma_{i+1}^+ \sigma_{i+2}^-\sigma_{i+3}^+ \sigma_{i+4}^-\sigma_{i+5}^+\,,
    \label{eq: qsi: EoM}
\end{align}
where the summation is over the six plaquettes that $i$ belongs to (see Fig.~\ref{fig:hexagons}) and $i+n$ denotes the $n$th spin counted from $i$ on each plaquette (the direction around the plaquette is immaterial). These equations were integrated numerically using the GNU Scientific Library implementation of the Prince--Dormand (8,9) ODE solver with automatic step size control~\cite{Galassi2009GNUManual}.

In order to generate thermally distributed classical spin configurations, we use a Monte Carlo algorithm that samples $\sigma^z$ and the phase of $\sigma^+$ (for its magnitude is fixed by $\sigma^z$) independently -- this mirrors the anisotropy of the Hamiltonian~\eqref{eq: qsi: ring exchange Hamiltonian}. Furthermore, we insist that there be no spinons in the system, that is, $\sum\sigma^z=0$ for all tetrahedra of the pyrochlore lattice. This can be achieved by updating $\sigma^z$ only in closed loops with alternating signs, similarly to typical low-temperature simulations of CSI~\cite{Barkema1998MonteModels,Melko2004MonteModel}. Unlike CSI, however, $\sigma^z$ is now a continuous variable, so loops can always be updated by small amounts. For convenience, we only perform updates around hexagons~%
\footnote[31]{
One can see that the hexagon updates lead to an ergodic Monte Carlo protocol within each $\sigma^z$ magnetisation sector as follows. The no-spinon constraint implies that $\sigma^z$ is the sum of a pure lattice curl and a global magnetisation (cf.~Appendix~\ref{appendix: interaction}). Therefore, for any two  valid configurations of $\sigma^z$ with equal overall magnetisation, there is an ``electric vector potential'' $G$ on the dual pyrochlore lattice such that $\curl G=\Delta\sigma^z$: Changing $\sigma^z$ around each hexagon by this $G$ on the corresponding dual pyrochlore site takes the system from one configuration to the other. In our work, we focus on the entropically dominant zero magnetisation sector only since none of the properties we look at depend on changes in the macroscopic magnetisation of the system. If needed, loop updates to sample different sectors could easily be introduced in the algorithm to remove this limitation.
}; 
the proposed change is drawn from a Gaussian distribution whose variance is proportional to temperature (this ensures a large  acceptance rate at all temperatures); and acceptance is decided using the Metropolis method.

On the other hand, there are no conservation laws to be obeyed by the phase $\phi$ of $\sigma^+$, so it can be updated independently for each spin. In particular, since the magnitude of $\sigma^+$ is kept constant, its distribution is given by 
\begin{equation}
    f(\phi_i)\propto e^{\beta \Re(\sigma_i^+ h_i^*)} = \exp \left[|\sigma_i^+|\,|h_i|\,\cos(\phi_i-\arg h_i)\right],
\end{equation}
where the effective exchange field $h_i$ is given by \eqref{eq: qsi: EoM}. This is a von Mises distribution which can be efficiently sampled directly~\cite{Best1979EfficientDistribution,Devroye1986Non-UniformGeneration}, that is, $\phi_i$ can be sampled without rejection. Furthermore, to eliminate spurious correlations due to not sampling the gauge freedom~\eqref{eq: qsi: gauge invariance}, each Monte Carlo step includes rotating $\phi$ for the four spins of each tetrahedron by an angle $\chi$ drawn from a uniform distribution.

It is important to note that, in the semiclassical picture, quantum correlations and uncertainties are irrelevant and therefore all quantities can be represented by pure numbers rather than quantum operators. That is, given a configuration of spins, the value of any observable can be determined unambiguously and straightforwardly. For intricate quantities such as the magnetic field $B$, this is a substantial improvement over standard methods, e.g., quantum Monte Carlo. 
In particular, we follow the soft spin prescription to identify $\sigma^\pm$ with $e^{\pm iA}$~\cite{Hermele2004PyrochloreMagnet} and take the vector potential $A$ to be the complex argument of $\sigma^+=\sigma^x + i\sigma^y$. Now, the magnetic field $B$ follows as
\begin{equation}
    B = \curl A = \arg(\sigma_1^+\sigma_2^-\sigma_3^+\sigma_4^-\sigma_5^+\sigma_6^-);
\end{equation}
the argument function is restricted to the interval $[-\pi,\pi)$ for a unique $B$ with the smallest possible modulus, consistently with Ref.~\cite{Hermele2004PyrochloreMagnet}. Furthermore, the choice of $\sigma_1$ for each plaquette affects the sign of $B$; this choice must be made with reference to the pyrochlore lattice geometry to obtain a self-consistent lattice vector field (see also Ref.~\cite{Benton2012SeeingIce} and Appendix~\ref{appendix: lattice fields}). 
%
%

\section{Photons}
\label{section: photon}

We demonstrate the validity and benchmark the accuracy of our method by investigating photon modes in the low-temperature dynamics of the system. Indeed, clean photon modes arise if their interaction with each other and with other excitations is minimised.
Spinons are  excluded altogether by the loop update Monte Carlo algorithm; visons and photon--photon interactions can be eliminated by reducing the temperature. 

\begin{figure}
    \centering
    \includegraphics{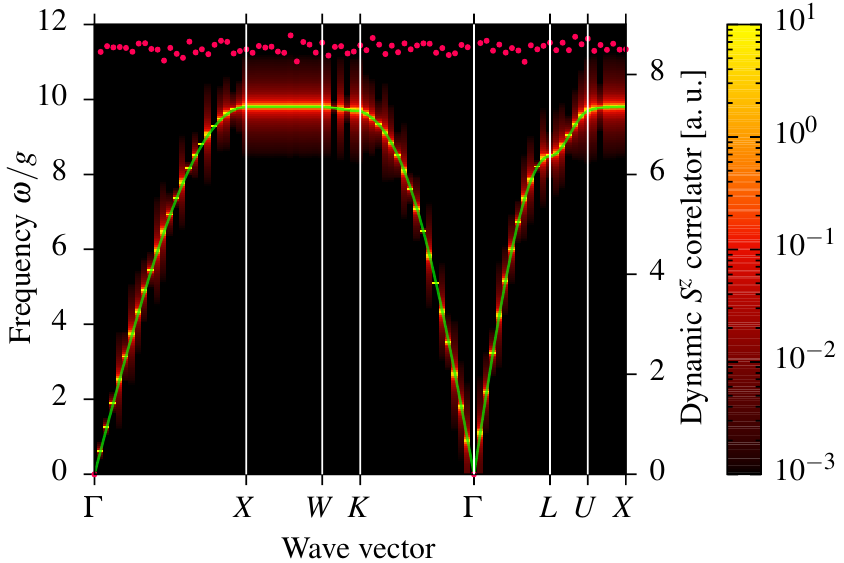}
    \caption{Dynamical structure factor $S^{zz}(\v q,\omega)$ along high symmetry directions in the semiclassical QSI model at $T=10^{-4}g$. Photons manifest as a sharp, gapless, linearly dispersing branch of classical normal modes. The frequency of these modes matches excellently with large-$S$ analytic predictions (green line)~\cite{Kwasigroch2017SemiclassicalIce,Note41}. The integrated structure factor of the modes (red dots) is independent of $\v q$, as expected on grounds of equipartition.
    }
    \label{fig:photon}
\end{figure}

We generated 4096 stochastically independent Monte Carlo configurations of a cubic sample of size $L=24$ unit cells at temperature $T = 10^{-4} g$ and calculated the time evolution of each for 2048 time steps of size $\delta t=(16g)^{-1}$. The results were Fourier transformed using the FFTW library~\cite{Frigo2005TheFFTW3} in time and space, separately for the four fcc sublattices of the pyrochlore lattice. Following Ref.~\cite{Huang2018DynamicsIce}, we evaluated the correlator
\begin{equation}
    S^{zz}(\v q, \omega) = \sum_\mu \big\langle \sigma^z_\mu(\v q,\omega)\sigma^z_\mu(-\v q,-\omega)\big\rangle 
\end{equation}
along high symmetry directions, where the summation runs over the sublattices; the results are plotted in Fig.~\ref{fig:photon}. A single set of remarkably sharp normal modes appear in the data~\footnote[41]{The period of oscillations is not necessarily commensurate with the simulation time window. After Fourier transforming, this leads to the slight broadening of the dispersion seen in Fig.~\ref{fig:photon}.}.
The frequencies of the numerically obtained modes match perfectly with analytic results for the large-$S$ photon dispersion~\cite{Kwasigroch2017SemiclassicalIce} (green line in Fig.~\ref{fig:photon}), confirming that the CSL simulated by our method is indeed equivalent to large-$S$ QSI.

We note that the integrated structure factor $S^{zz}(\v q) = \int\!\ud\omega\, S^{zz}(\v q,\omega)$ is independent of $\v q$. This contrasts spin-1/2 QSI where $S^{zz}(\v q)\propto\omega(\v q)$~\cite{Benton2012SeeingIce}. This discrepancy is caused by the different realisation of photons in the two systems. In spin-1/2 QSI, zero point fluctuations of photon modes of energy $\hbar\omega(\v q)/2$ give rise to dynamic correlators even at zero temperature; in the large-$S$ classical spin liquid, photons are classical normal modes with energy determined by equipartition, leading to $\v q$-independent correlators~\footnote[49]{In terms of a large-$S$ expansion, we have set the magnitude of our spins, $\hbar S$, to 1. Therefore, $\hbar\sim1/S$ and thus the energy of individual photons, $\hbar\omega(\v q)$, also scales as $1/S$. In our simulations, $S\to\infty$ and so $T\gg\hbar\omega$ at any nonzero temperature. Photon populations are thus always large and can be thought of as being in a coherent, classical state.}. 
%
%

\section{Energetics of bare visons}
\label{section: bare vison}

$U(1)$ visons in QSI are $2\pi$-quantised sources of the emergent magnetic field $B$. Their existence and quantisation is due to the $2\pi$-ambiguity of the transverse phases $A$ that are promoted to a vector potential in the gauge theory description. For the same reason, however, specifying vison numbers unambiguously is far from trivial. Typically, visons are understood through their far-field effects, where $B\sim1/r^2$ is small and thus well defined, and the total flux across a large, closed surface gives a unique vison charge~\cite{Savary2017QuantumReview}. In principle, one can define a $U(1)$ vison charge operator 
\begin{equation}
    q = \div B/2\pi
    \label{eq: bare vison: charge operator}
\end{equation}
for each dual diamond site by giving $B$ on each plaquette a unique value (e.g., by restricting it to between $-\pi$ and $\pi$). However, since visons are dynamical, the system will not normally be in an eigenstate of $q$, which makes pinpointing visons complicated.
A great advantage of the semiclassical method is that observables like $B$ and $q$ are pure numbers rather than quantum operators: This means that the vison charge as defined above is always an unambiguous integer for all dual diamond sites. 

The semiclassical simulation also allows us to introduce visons at will. We achieve this via a fundamental step that consists of rotating the transverse components of the six spins around a given plaquette by $\pi/4$ in alternating directions. Doing so changes $B$ on the chosen plaquette by $-3\pi/2$ and by $\pi/2$ on its neighbours. Assuming that $B$ was small initially, one can regard the former as a change of $B$ by $\pi/2$ together with a $2\pi$ phase slip that inserts a vison pair across the plaquette in question. That is, this elementary step inserts a pair of visons with a symmetric near-field pattern around them. Repeating it along a chain of sites amounts to moving the visons apart from one another, similarly to the insertion of spinons in classical spin ice. 

The state generated by these local operations, however, is not the least energetic one, for they do not capture the change in $B$-field away from the visons or the ``Dirac string'' connecting them. While it may be possible to construct operators acting on all spins that have a larger overlap with the ``true vison creation operator''~\cite{Hermele2004PyrochloreMagnet}, we adopt here a simpler and more straightforwardly reliable approach. We equilibrate the photons generated by the local vison creation operation using the Monte Carlo algorithm and gradually reduce the effective temperature until the remaining photon population can be ignored, leaving behind a two-vison metastable state. The only issue of this method is vison movement: Visons in the semiclassical model are not inherently mobile~\cite{Michal} (there are no explicit vison hopping terms in the Hamiltonian), but a high-temperature photon cloud can move them around. We find that starting the annealing procedure at a sufficiently low temperature prevents such motion unless the visons are introduced within a distance of about $2a_0$ (where $a_0$ is the fcc lattice parameter). 

\begin{figure}
    \centering
    \includegraphics{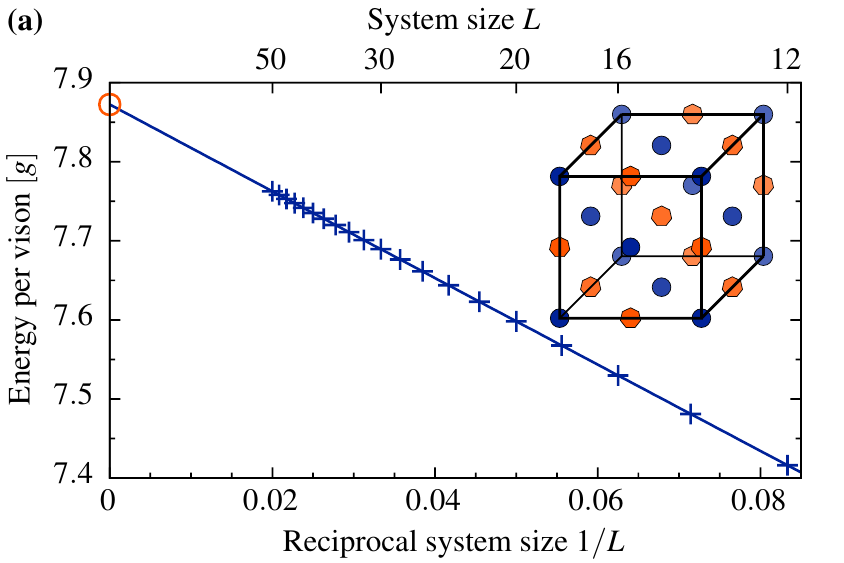}
    \includegraphics{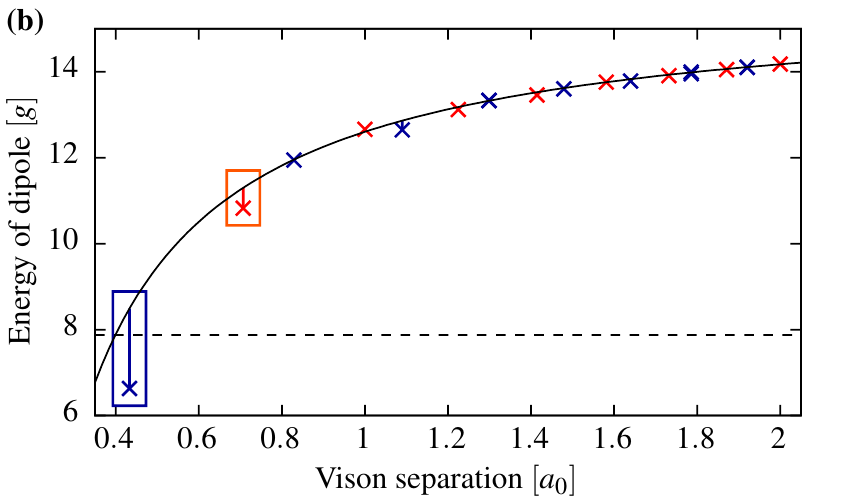}
    \caption{\textbf{(a)} Energy per vison in a rock-salt arrangement of visons (see inset), as a function of linear system size $L$. In a Coulomb interacting system, this energy is linear in $1/L$, see~\eqref{eq: bare vison: rock salt energy function}; the data indeed follow this trend excellently. The energy of an isolated vison is given by extrapolating to $L=\infty$ (red circle); the Coulomb interaction strength can be extracted from the slope of the fitted straight line. 
    \textbf{(b)} Energy of a pair of nearby positive and negative visons as a function of separation. Energy estimates based on the full semiclassical theory match the Coulomb law $\varepsilon(r) = 2\mu-\alpha/r$ (solid line) well for all but nearest and second nearest neighbours. In particular, notice that the energy cost of a nearest neighbour dipole is smaller than that of a single vison (dashed line).}
    \label{fig:freeze}
\end{figure}

The energy difference between the resulting configuration and the ground state can be regarded as the energy cost of two visons plus their Ewald summed interaction energy. The latter, however, contains a surface term~\cite{deLeeuw1980SimulationConstants} which makes interpreting the results complicated, see Appendix~\ref{appendix: interaction: ewald surface}. 
This can be remedied by using an arrangement of visons with no net dipole moment, such as the rock salt configuration in Fig.~\ref{fig:freeze}(a). This arrangement was set up using the local vison insertion protocol described above; photons were equilibrated for $16L$ Monte Carlo steps~%
\footnote[51]{A ``Monte Carlo step'' in this paper consists of the following: sampling the $xy$ phase angle $\phi$ of each spin; a Metropolis attempt to change $\sigma^z$ around each hexagonal plaquette; and sampling the gauge freedom of $\phi$ on each tetrahedron. These elementary steps are described in more detail in Sec.~\ref{section: qsi}.}
at $T=g/256$; then cooled 32 times by a factor of 2 and equilibrated for $8L$ steps each time. In the end, the temperature of the photon cloud reached $2^{-40}g\approx 9\cdot10^{-13}g$;  equilibration was monitored through the acceptance rate of Metropolis steps which remained consistently high throughout the process. The so obtained metastable energies are plotted in Fig.~\ref{fig:freeze}(a).

Modelling visons as Coulomb interacting charges, the energy of the rock salt configuration is 
\begin{equation}
    \varepsilon(L) = \mu - \frac{M\alpha}{L a_0}
    \label{eq: bare vison: rock salt energy function}
\end{equation}
per vison, where $\mu$ is the energy cost of an isolated vison, $L$ is the size of the cubic simulation box in units of the fcc lattice parameter $a_0$, $\alpha$ is the Coulomb interaction strength, and $M=1.74756459\dots$ is the appropriate Madelung constant. Fitting this form to the numerical results plotted in Fig.~\ref{fig:freeze}(a) yields
\begin{align}
    \mu &= 7.872367608(68)\, g; &
    \alpha &= 3.1416145(37)\, ga_0\,.
    \label{eq: bare vison: mass, interaction}
\end{align}

The numerical results can be compared to an analytic estimate of the energy cost and interaction strength of visons from a quadratic approximation to the energy of the magnetic field, $-g\cos B$, see Appendix~\ref{appendix: interaction: coulomb}. The quadratic estimate of $\mu$, $8.858g$, is substantially different from the numerical result~\eqref{eq: bare vison: mass, interaction}; however, the interaction strength, $\alpha=\pi ga_0$, matches excellently. This is so because the vison energy cost includes that of its immediate neighbourhood, where the quadratic theory breaks down; on the contrary, most of the interaction energy is due to the overlap of far fields for which the quadratic theory is accurate.

\begin{figure*}
    \centering
    \includegraphics{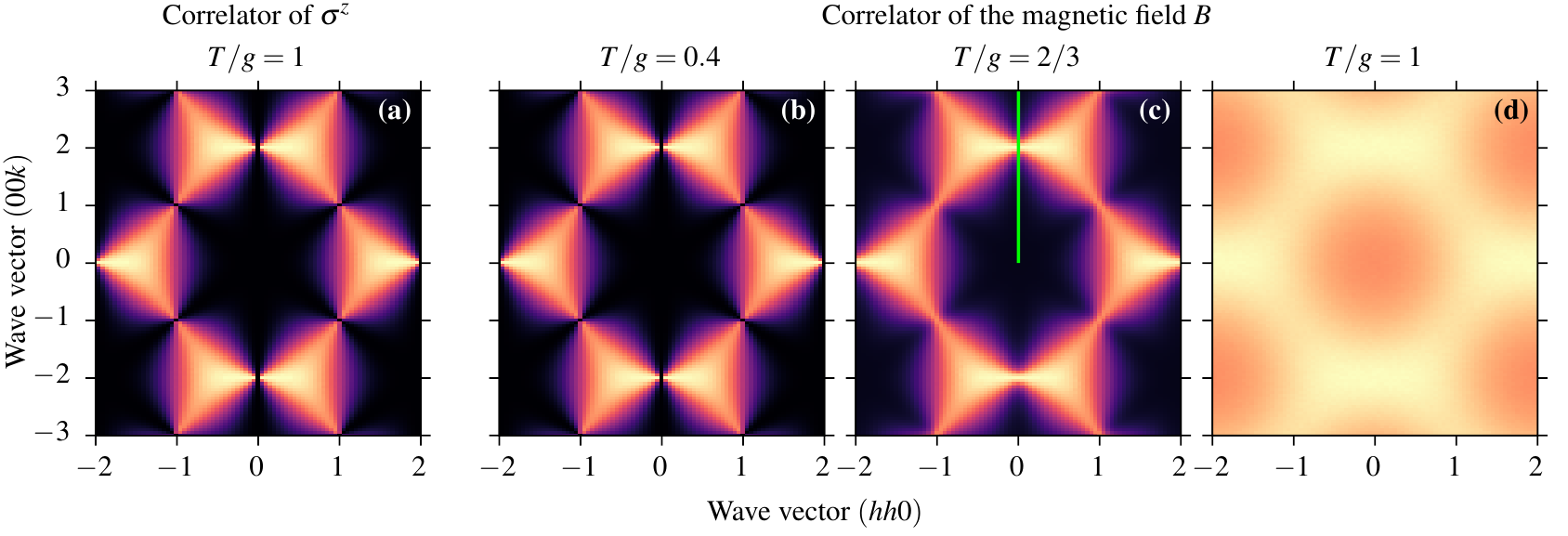}
    \caption{Static correlation functions $\langle \sigma^z(-\v q)\sigma^z(\v q)\rangle$ \textbf{(a)} and $\langle B(-\v q)B(\v q)\rangle$ \textbf{(b--d)} in the semiclassical QSI model. Due to the no-spinon constraint enforced in our Monte Carlo algorithm, the former retains sharp pinch points at all temperatures. Unlike spin-1/2 QSI~\cite{Shannon2012QuantumStudy,Benton2012SeeingIce}, these pinch points are not depleted near the $\Gamma$ points. At low temperatures, the emergent magnetic field gives rise to identical pinch points which are blurred by thermally induced visons  at higher temperatures; at $T\gtrsim g$, the pinch point pattern is washed out altogether.}
    \label{fig:pinch_points}
\end{figure*}

We finally consider how accurately the interaction of nearby visons is described by the asymptotic Coulomb law. 
Since these visons were unstable against the photon cooling protocol, an alternative technique had to be used.
Within quadratic approximation, the $B$-field configuration that minimises the magnetic energy for a given arrangement of visons can be found explicitly, as described in Appendix~\ref{appendix: interaction: coulomb}: The sum of $-g\cos B$ for all plaquettes in this configuration is an upper bound on the true energy of the vison configuration. 
We benchmarked this estimate against the photon cooling technique at vison separations where the latter is applicable; because of the good agreement there, we decided to use this method to estimate the energy of vison dipoles with separations smaller than $2a_0$. The results are plotted in Fig.~\ref{fig:freeze}(b) together with the Coulomb energy estimate $2\mu-\alpha/r$, where $\mu$ and $\alpha$ are taken from~\eqref{eq: bare vison: mass, interaction}. The latter is a remarkably good approximation even at third nearest neighbour distance; there is a discrepancy of about $0.5g$ for next-nearest neighbours and about $2g$ for nearest neighbours. 

Most notably, the energy of a nearest neighbour dipole is smaller than that of a single isolated vison, $\mu$. Visons in QSI thus form a weak electrolyte. While deconfined, their dissociation is so energetically unfavourable that most visons at low temperatures remain associated with an oppositely charged one. This behaviour is quite different from that of spinons in CSI, whose energy cost is set independently by the dominant Ising exchange interaction, and they remain energetically dissociated even at nearest-neighbour distance in the presence of entropic and dipolar interactions.
%
%

\section{Thermodynamics of photons and visons}
\label{section: thermo}

A natural way to introduce gapped excitations in numerical simulations is via thermal fluctuations at finite temperature where an equilibrium population of such excitations arises. Thermodynamic quantities like heat capacity or thermal conductivity are promising signatures of gapped quasiparticles and QSL behaviour in general~\cite{Knolle2019ALiquids,Rau2019FrustratedPyrochlores,Tokiwa2018DiscoveryLiquid}.
Therefore, we studied the interactions of visons and photonic modes in a finite temperature ensemble. This was greatly aided by the ability of our method to directly access observables such as the emergent magnetic field and the vison charge operator~\eqref{eq: bare vison: charge operator}.
%
%

\subsection{Magnetic field pinch points}

We evaluated static correlation functions of the emergent magnetic field, $\langle B(-\v q)B(\v q)\rangle$, at temperatures between $0.4g$ and $1g$ using Monte Carlo spin configurations of a cubic sample of size $L=20$ unit cells~%
\footnote[61]{55 temperature points were used, uniformly distributed in $1/T$. For temperatures above $0.5g$, 131\,072 stochastically independent Monte Carlo samples were generated; for those between $0.4g$ and $0.5g$, 262\,144 samples were used.}. 
These correlators are plotted in the $(hhk)$ plane for three temperatures in Fig.~\ref{fig:pinch_points}; the correlators of the emergent electric field $\sigma^z$ are also shown for comparison. At low temperatures, both correlators exhibit sharp pinch points at the $\Gamma$ points in the pattern familiar from classical spin ice~\cite{Isakov2004DipolarMagnets,Henley2005Power-lawAntiferromagnets}. The pinch points remain sharp for the electric field at all temperatures as spinons are excluded by the Monte Carlo algorithm. On the contrary, the introduction of visons blurs the $B$-field pinch points in much the same way monopoles blur CSI pinch points~\cite{Fennell2009MagneticHo2Ti2O7,Sen2013CoulombDisorder}.
This picture is in some departure from spin-$1/2$ QSI in which pinch points are suppressed near the $\Gamma$ points and thus no sharp features appear~\cite{Shannon2012QuantumStudy,Benton2012SeeingIce}. This is due to the different way in which photons appear in the two systems. In the semiclassical case, they are classical normal modes, hence their energy content is constant as per equipartition~\cite{Note49}; in the spin-$1/2$ case, low-temperature physics is dominated by quantum zero-point fluctuations of photon modes which give rise to correlators proportional to the photon dispersion $\omega(\v q)$, thus suppressing the pinch points.

\begin{figure}
    \centering
    \includegraphics{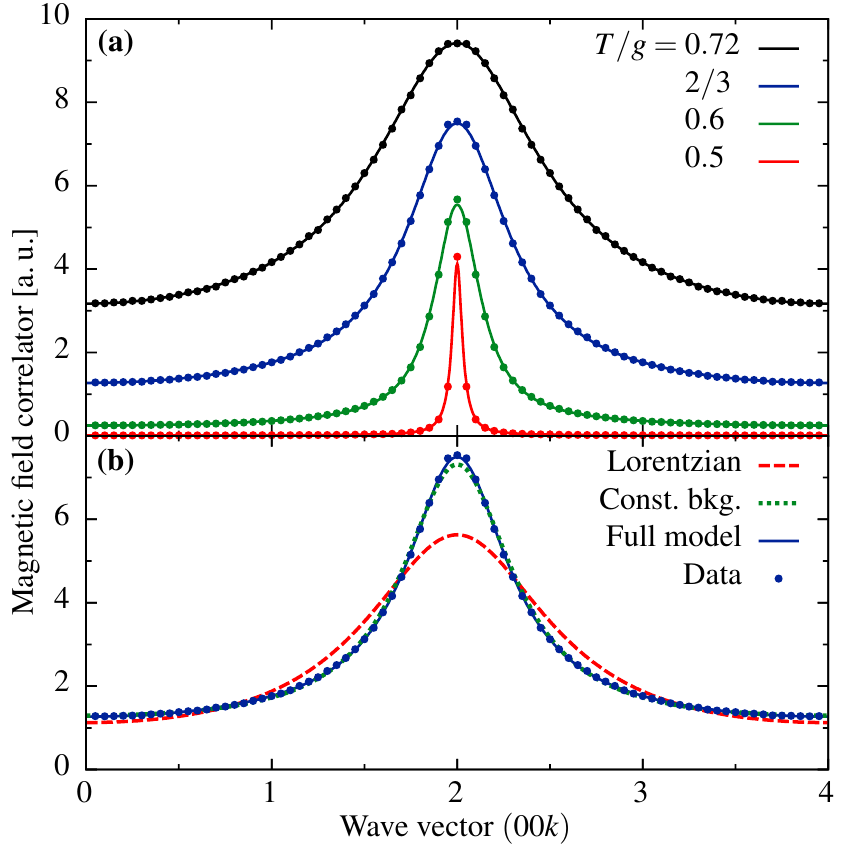}
    \caption{\textbf{(a)} Static correlation function of the emergent magnetic field, $\langle B(-\v q)B(\v q)\rangle$, along the $(00k)$ axis for four different temperatures in the semiclassical QSI model. All data  sets (coloured dots) fit excellently to the theoretical form~\eqref{eq: thermo: pinch point blurring theory} (solid lines). 
    \textbf{(b)} Simulation data at $T/g=2/3$ (blue dots) fitted to several theoretical models. A simple Lorentzian (red dashed line), corresponding to a Debye plasma of visons, provides a poor fit.  An additional constant background, corresponding to nearest-neighbour vison dipoles, improves this fit substantially (green dotted line). A further minor, but significant, improvement can be achieved by including a contribution for next-nearest-neighbour vison pairs~\cite{Note62}, giving rise to the form~\eqref{eq: thermo: pinch point blurring theory} (blue line).}
    \label{fig:pinch_fit}
\end{figure}

Pinch point blurring is a common experimental diagnostic of spinons in CSI~\cite{Fennell2009MagneticHo2Ti2O7}; likewise, we were able to extract quantitative information about the visons from the blurring of $B$-field pinch points. We focus on the $(00k)$ axis (green line in Fig.~\ref{fig:pinch_points}), where the photon contribution to the correlator vanishes~\cite{Isakov2004DipolarMagnets,Henley2005Power-lawAntiferromagnets}, that is, the signal is entirely due to visons. Cuts of the correlator along this axis are plotted in Fig.~\ref{fig:pinch_fit}(a) for four temperatures. These show an apparently Lorentzian peak at the $(002)$ pinch point, indicating a Debye plasma of visons. 
Unlike pinch point blurring in CSI, however, our blurring pattern is not explained by a Lorentzian peak in itself; in particular, the peak appears on top of a substantial constant background, as demonstrated in Fig.~\ref{fig:pinch_fit}(b). This constant correlator can be ascribed to a large population of nearest neighbour vison dipoles that are independent of the Debye plasma mentioned above; for a detailed derivation, see Appendix~\ref{appendix: pinch point}. This underlines the observation that visons in QSI form a weak electrolyte, that is, they interact strongly enough that a large fraction of their thermal population remains associated, as discussed in Sec.~\ref{section: bare vison}.

Furthermore, the Coulomb energy formula for vison pairs substantially overestimates the energy cost of next-nearest-neighbour vison dipoles, see Fig.~\ref{fig:freeze}(b). Therefore,  we anticipate an excess population of them compared to that predicted by the Debye--H\"uckel approximation. While this effect is not qualitative, it does introduce a correction to the $B$-field correlator that is proportional to $\sin^2(q_z/8)$ along the $(00k)$ axis, see Appendix~\ref{appendix: pinch point: dipoles}. 
To take into account the effects of these closely associated dipoles, we fitted the functional form
\begin{equation}
    \langle B(-\v q)B(\v q)\rangle = C_1 + C_2 \sin^2(q_z/8) + \frac{C}{1+\big[8\xi\cos(q_z/8) \big]^2}
    \label{eq: thermo: pinch point blurring theory}
\end{equation}
to the data at all temperature points; $\xi\propto\rho_\mathrm{free}^{-1/2}$ is the Debye length of the plasma formed by dissociated dipoles, where $\rho_\mathrm{free}$ is the density of dissociated visons; while $C_1$ is proportional to the density of nearest neighbour vison pairs, $\rho_\mathrm{nn}$~\footnote[62]{The complicated form of the Lorentzian is to account for the periodicity of the data imposed by the lattice; $C_2$ is related to the \textit{excess} density of next-nearest-neighbour dipoles, a quantity that is hard to get a direct handle on}. Equation~\eqref{eq: thermo: pinch point blurring theory} fits the data excellently throughout the investigated temperature range; Fig.~\ref{fig:pinch_fit}(b) demonstrates that all three terms are necessary to achieve this, although a very good agreement is already obtained without the $C_2$ contribution. 
%
%

\subsection{Temperature dependence of vison density}

\begin{figure}
    \centering
    \includegraphics{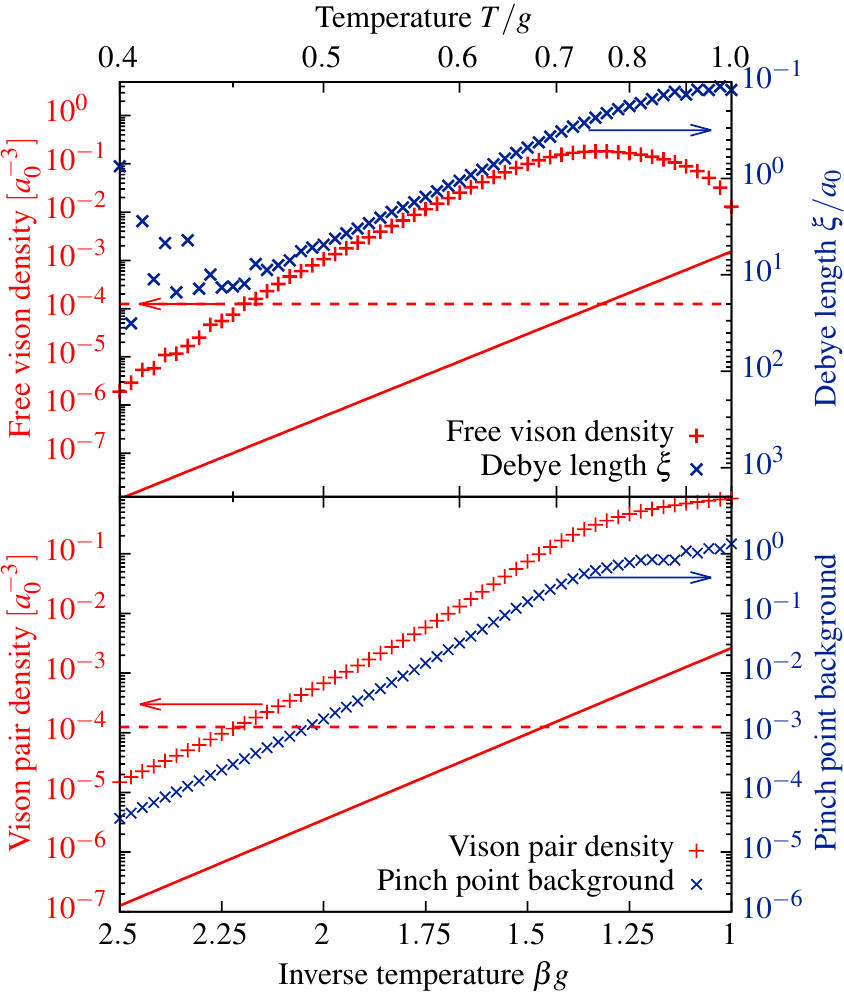}
    \caption{Arrhenius plots of the density of dissociated visons $\rho_\mathrm{free}$ (top panel) and nearest-neighbour vison dipoles $\rho_\mathrm{nn}$ (bottom panel) and of the Debye length $\xi$ (top panel) and background correlator $C_1$ (bottom panel) extracted from pinch point blurring data. The logarithmic scales of the left and right axes are linked to each other by the theoretical relations $\xi\propto\rho_\mathrm{free}^{-1/2}$ and $C_1\propto \rho_\mathrm{nn}$, with an arbitrary scaling offset; the simulation data indeed follow these relations, even below the limit of a single vison (pair) in the entire system (red dashed lines). The densities of both vison populations follow an approximate Arrhenius relation with gaps similar to zero-temperature ones; their values, however, are much larger than a simple Boltzmann factor prediction (red solid lines).} 
    \label{fig:pinch_count}
\end{figure}

Visons can also be counted explicitly in our Monte Carlo simulations. Besides the number $N$ of all visons, the number $N_\mathrm{nn}$ of dual diamond lattice bonds with two visons of opposite charge on their ends was obtained; from these, the number of dissociated visons was estimated as $N_\mathrm{free}=N-2N_\mathrm{nn}$. The density of bound dipoles and free visons was plotted as a function of inverse temperature in Fig.~\ref{fig:pinch_count}, together with the constant background $C_1$ of the $B$-field correlator and the Debye length $\xi$, respectively; we expect $\xi\propto\rho_\mathrm{free}^{-1/2}$~\cite{Levin2002ElectrostaticBiology} and $C_1\propto \rho_\mathrm{nn}$ (see Appendix~\ref{appendix: pinch point: dipoles}). These relations hold quite well throughout the temperature range, confirming that magnetic field pinch point blurring is a good measure of  vison populations. 

The densities of both dissociated and bound visons follow an approximate Arrhenius law at low temperatures with an apparent gap close to the bare vison energy~\eqref{eq: bare vison: mass, interaction} and the nearest neighbour dipole energy shown in Fig.~\ref{fig:freeze}(b), respectively. However, these vison densities saturate at $T\approx g$, an order of magnitude below the bare energy costs, and correspondingly, densities at $T\lesssim g$ are much larger than predicted by a simple Boltzmann factor with quasi-equilibrium gaps (red lines in Fig.~\ref{fig:pinch_count}). %
%

\subsection{Thermodynamic effects of quasiparticle interactions}

\begin{figure}
    \centering
    \includegraphics{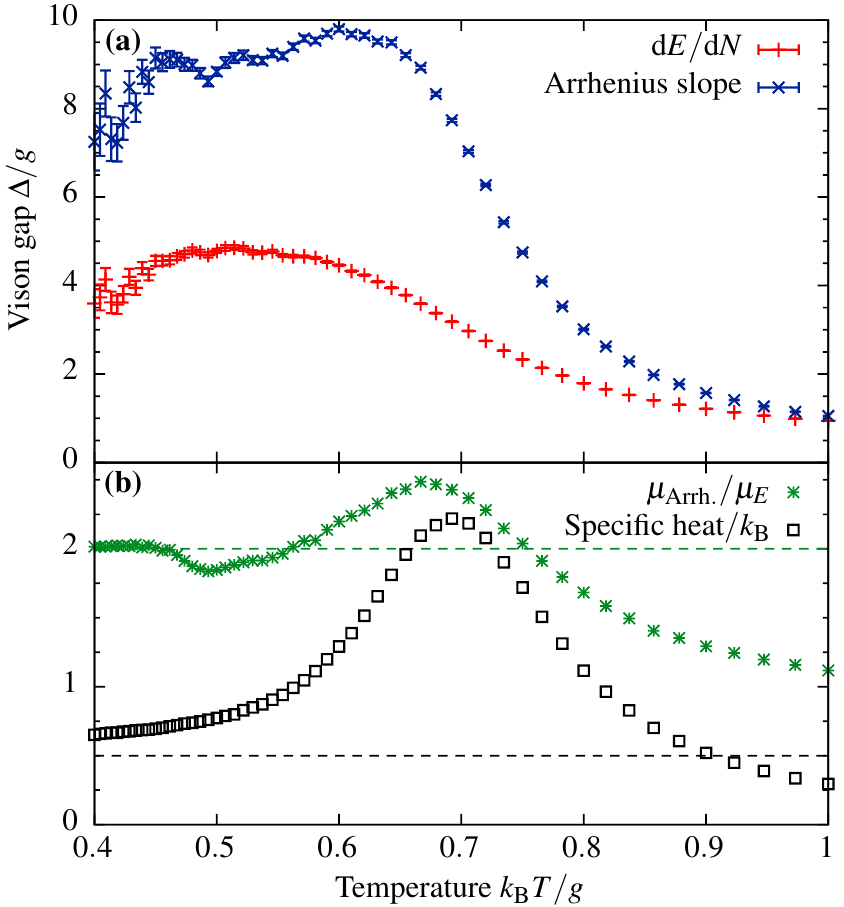}
    \caption{\textbf{(a)} Effective energy cost $\mu_E = \ud\langle E\rangle/\ud N$ (red) and effective Arrhenius gap $\mu_\mathrm{Arrh.}=-\ud\log N/\ud\beta$ (blue) of visons as a function of temperature. The latter is somewhat above the zero-temperature energy cost of nearest-neighbour vison dipoles at low temperatures. At $T\gtrsim0.7g$, both gap estimates decline steeply. 
    \textbf{(b)} Ratio of the gap estimates, $\mu_\mathrm{Arrh.}/\mu_E$ (green stars), and specific heat capacity of the system per spin (black squares) as a function of temperature (in units of $k_\mathrm{B}$ per spin). The former tends to 2 at low temperatures (red dashed line), corresponding to nearest-neighbour vison pairs as the dominant low-temperature vison population; at intermediate temperatures, larger coherent vison clusters raise it further. At low temperatures, the specific heat capacity tends to $k_\mathrm{B}/2$ (blue dashed line) due to equipartition of photon modes; it then increases due to photon interactions, turning into a vison Schottky peak at $T\approx0.7g$; beyond that, the specific heat drops below the photon equipartition limit, indicating the breakdown of photons as quasiparticles.    }
    \label{fig:gap}
\end{figure}

In order to identify the origin of this excess density of visons, we considered two different estimates of their energy cost in the thermal ensemble: the dependence of energy on vison number, quantified by $\ud \langle E\rangle /\ud N$, and the slope of the Arrhenius plot $\log N$ versus $1/T$. 
Both quantities were evaluated using data recorded at single temperature points, similarly to heat capacity estimation using the fluctuation--dissipation theorem~\cite{Landau5}:
\begin{subequations}
\begin{align}
    \mu_E &= \left.\frac{\ud \langle E\rangle}{\ud N}\right|_T = \frac{\cov(E,N)}{\var N}\\
    \mu_\mathrm{Arrh.} &=-\frac{\ud\log N}{\ud\beta} = \frac{\cov (E,N)}{N}\,;
\end{align}
\end{subequations}
for a derivation, see Appendix~\ref{appendix: gap estimate}.  These estimates are plotted in Fig.~\ref{fig:gap}, together with $\mu_\mathrm{Arrh.}/\mu_E =\var N/N$. At low temperatures, the distribution of $N$ is generated by a Poisson distribution of thermal excitations which may well be a collective one made up of several visons: It follows (see Appendix~\ref{appendix: gap estimate}) that $\var N/N$ gives the typical vison content of such a collective excitation.

At the lowest temperatures, $\var N/N$ tends to 2; this again demonstrates that the dominant vison species at low temperatures are bound dipoles. At intermediate temperatures, $\var N/N$ increases further, to about 2.5, indicating collective excitations consisting of more than two visons. Salient examples of such  clusters could be `polarons' consisting of a free vison and nearest-neighbour dipoles. As the $B$-field emanating from the former is quite large in its immediate neighbourhood, the energy cost of appropriately aligned dipoles decreases substantially, causing them to proliferate. An effective phenomenological model based on this mechanism that allows us to gain a qualitative understanding is presented in Appendix~\ref{appendix: toy partition function}. At even higher temperatures, $\var N/N$ decreases as $N$ approaches saturation.

We also note that the energy cost of visons in the thermal ensemble declines steeply for $T\gtrsim 0.7g$. This is due to interactions between visons and photons arising from corrections to the gauge theory beyond quadratic order. The next order in the Villain expansion~\cite{Kwasigroch2017SemiclassicalIce} consists of negative quartic terms which, in the presence of a (thermal) population of photons, lead to a negative renormalisation of vison energies and interactions; see Appendix~\ref{appendix: toy partition function} for a more detailed discussion of both the mechanism and its effect on the effective vison gap. This mechanism is at work at intermediate temperatures, reducing the gap of visons compared to the zero-temperature value, resulting in larger thermal populations. This may also explain the excess vison density observed in Fig.~\ref{fig:pinch_count}.

As temperature is increased, the energy cost of visons drops substantially and their density saturates. Eventually, at temperatures $T\gg g$, the visons cease to be useful quasiparticles as it is to be expected from the perspective of the individual spins:
The contribution of each hexagonal plaquette to the Hamiltonian~\eqref{eq: qsi: ring exchange Hamiltonian} is $O(g)$, therefore, all spin configurations satisfying the no-spinon constraint (which is still enforced by the Monte Carlo algorithm) become roughly equally likely, regardless of vison content. The situation is similar to the crossover of classical spin ice into a high-temperature paramagnetic phase as spinons cease to be useful quasiparticles to describe the system. 
In our case, visons and photons are washed out at high temperatures, giving rise to a classical spin ice phase with the spinon as its only excitation. Indeed, the heat capacity of the system, plotted in Fig.~\ref{fig:gap}, displays a Schottky peak at $T\approx0.7 g$ above which it drops below the equipartition heat capacity of photon modes, $k_\mathrm{B}/2$ per spin, and tends to zero as $T\to\infty$. This indicates that photons break down as quasiparticles together with visons. 
%
%

\section{Conclusion}
\label{section: conclusion}

We developed a semiclassical numerical technique to simulate quantum spin liquids whose physics is underpinned by perturbative ring exchange processes. 
Thus far, these systems remained elusive to classical simulations, as their native bilinear exchange Hamiltonians tend to predict ordered or Ising spin liquid phases, devoid of QSL behaviour. In contrast, we studied effective ring-exchange Hamiltonians directly, which are expected to remain in the quantum spin liquid phase in the large-$S$ limit. We formulated our method in terms of a large-$S$ path integral formalism and demonstrated that a combination of classical Monte Carlo sampling and Landau--Lifshitz dynamical evolution captures the least action trajectories of the finite-temperature path integral of the system.
Treating spins semiclassically allowed us to access a variety of observables not readily available to other simulation methods. 
We thus gained a detailed insight into the behaviour of quasiparticles in these quantum spin liquids, including gapped excitations which are not yet amenable to other computational and analytic techniques. 

We demonstrated the potential of our method on pyrochlore quantum spin ice, a paradigmatic $U(1)$ QSL~\cite{Hermele2004PyrochloreMagnet}. We identified a gapless, linearly dispersing branch of classical normal modes in spin dynamics. At low temperatures, these photonic modes are remarkably sharp and their dispersion matches analytic results from large-$S$ path integral calculations extremely well~\cite{Kwasigroch2017SemiclassicalIce}. We also observed gapped, quantised vortices of the emergent magnetic field with the phenomenology expected for the elusive $U(1)$ vison quasiparticles of QSI. In a showcase of the capabilities of our method, we were able to introduce these visons in a controlled way in the system and study their interactions in vacuum (that is, in an effectively zero-temperature photon background). 
In contrast with the behaviour of spinons, the energy cost and interaction strength of visons are set by a single parameter in the original Hamiltonian, that is, they are not separated parametrically. We found that the interaction of a nearest-neighbour pair of visons is strong enough that the energy cost of the resulting dipole is less than that of an isolated vison; namely, the visons form a weak electrolyte. At low temperatures, this leads to a relatively large population of closely associated visons which must be taken into account in modelling and understanding their behaviour, in particular in effective descriptions of vison dynamics~\cite{Chen2016MagneticYb2Ti2O7,Chen2017DiracsClassification, Michal}. 

We also studied the interplay of photonic modes and visons in thermal equilibrium at finite temperature. We developed a detailed understanding of visons through directly tracking them and studying their effect on pinch point patterns of the relevant static correlation functions. Similarly to the well understood case of spinons in classical spin ice, these pinch points blur due to the presence of visons; by contrast, however, they are also uniformly lifted by a population of associated visons. These effects can be used to reliably measure the density of associated and dissociated visons separately. We find that they are thermally activated with effective gaps comparable to their respective energy costs in the zero-temperature limit. 

A simple Boltzmann factor calculation, however, substantially underestimates vison populations at intermediate temperatures, which also saturate at temperatures far smaller than the $T\to 0$ vison cost. 
We understand this to be a consequence of strong interactions between visons and a highly excited photon background. The latter introduces fluctuations in the emergent electric and magnetic fields that blur visons, reducing their energy and leading to larger populations than naively expected. 
In spin-1/2 QSI, vacuum fluctuations of the photon modes occur even at zero temperature; we anticipate that these would similarly reduce the chemical potential of visons, possibly bringing it within the photon band. Since vison--photon interactions are controlled by the same ring-exchange energy scale as the energy of both quasiparticles, visons and photons of similar energy may then readily hybridise, leading to composite particles with more elaborate and exciting features than those of either visons or photons. 
It would be interesting to further test and confirm this behaviour in a spin-1/2 quantum setting, e.g., using quantum Monte Carlo. In particular, our work suggests that static correlators of the emergent magnetic field, possibly cumbersome but likely accessible to QMC, present a promising angle to study visons.
Such studies would also be instrumental in confirming the weak electrolyte behaviour found in the semiclassical limit.

One direct experimental signature that was identified in our work is the Schottky peak in the specific heat due to visons (see Fig.~\ref{fig:gap}). However, one must keep in mind that quantum photon excitations also contribute a Debye term to the specific heat, which is absent in semiclassical photonic normal modes. Hybridisation between photons and visons may well merge the two contributions and alter the shape of the Schottky anomaly.

The semiclassical numerical technique can naturally be applied to the dynamics of a high-temperature ensemble that contains thermally generated visons. However, the dynamics of visons is far from trivial, since they are immobile at zero temperature~\cite{Michal} and their motion at finite temperature is due to being ``tugged'' by the photon background. Further work is needed to gain better insight into this behaviour. 

A particularly interesting experimentally relevant direction is understanding magnetic noise in QSI~\cite{Dusad2019MagneticNoise}: Here, our method could be readily used to investigate the low temperature regime where the leading magnetic noise contribution is due to photons and spinon excitations can be neglected. 
Similarly, our approach could be applied to better understand the photon and vison contribution -- and their interplay, in particular the weak electrolyte behaviour -- to thermal transport in candidate QSI materials~\cite{Tokiwa2018DiscoveryLiquid}.

Whereas the gauge charge and fields due to (static) visons are not directly accessible in experiments, it is often the case that gauge degrees of freedom also carry irrational physical ones~\cite{Moessner2010IrrationalOrder}.
For example, magnetostriction or Dzyaloshinskii--Moriya effects can associate real electric polarisation with the emergent gauge fields of QSI~\cite{Khomskii2012ElectricIce,Lantagne-Hurtubise2017ElectricIce,Nakosai2019MagneticJunction}. As a result, one may expect, for instance, the vison pairs to have a signature in the dielectric response and electric susceptibility of QSI materials.

Our method may be extended to include terms in the simulated Hamiltonian that enable introducing static or dynamical spinons. Since spinons appear to have salient experimental signatures  in, for instance, magnetisation and neutron scattering measurements~\cite{Huang2018DynamicsIce}, a better understanding of spinon--vison interactions through our simulations may provide experimentally accessible handles to study collective photon and vison behaviour. 
It is important to note, however, that spinons are not quantised in the semiclassical setting (see Sec.~\ref{section: simulation}); therefore, understanding what their behaviour tells about the original quantum problem requires some care. 

Beyond QSI, our approach is manifestly applicable to other quantum spin liquids underpinned by ring exchange processes. Quantum dimer models are a salient example, especially in light of novel large-$S$ analytical approaches~\cite{Szabo2019GeneralisedConstraints} that apply also to non-bipartite lattices and their $\mathbb{Z}_2$ RVB phases~\cite{Moessner2001ResonatingModel}. Our technique could add to the understanding of these systems, in particular to the interplay of the $\mathbb{Z}_2$ vison excitations with the dimer liquid background as correlations develop, e.g., upon approaching quantum critical points out of the $\mathbb{Z}_2$ RVB phase.

Finally, the ability of the method to naturally include quantised charges [namely, U(1) visons] in an effectively classical system may have ramifications for lattice gauge theories in general. 
Vison excitations in our semiclassical model are quantised solitons, similarly to Dirac monopoles in quantum electrodynamics, and like these monopoles, they are likely to be in a strong coupling regime. The reduction of vison energy by a thermally fluctuating background may indeed be a semiclassical analogue of running couplings that are brought about by virtual particle--antiparticle bubbles in QED. 
This is but a speculative yet intriguing potential connection between QSI and the strong coupling problem in QED, which warrants further investigation in future work. 
%
%

\section*{Acknowledgements}

We thank Cristian Batista, John Chalker, Felix Flicker, Roderich Moessner, Nic Shannon, and David Tong for stimulating discussions. We are particularly grateful to David Ho for bringing the idea of a possible analogy to the strong coupling problem in QED monopole production to our attention, and to Micha\l\ Kwasigroch for sharing preliminary results of his work~\cite{Michal} and for engaging in fruitful conversations with us on numerous occasions. 
Fig.~\ref{fig:hexagons} was prepared using VESTA 3~\cite{Momma2011VESTAData}. Figs.~\ref{fig:photon}, \ref{fig:pinch_points}, and~\ref{fig:nnn_correlators} use perceptually uniform color maps developed in Ref.~\cite{Kovesi2015GoodThem}.
This work was supported in part by Engineering and Physical Sciences Research Council (EPSRC) Grants No. EP/P034616/1, EP/M007065/1, and EP/K028960/1.
%
%
\appendix

\section{\label{appendix: lattice fields}
Scalar and vector fields on the diamond lattice}

Lattice vector fields can be defined on the bonds of any bipartite lattice; in order to imbue them with a sense of directionality, their sign is taken to depend on the orientation $\v r\to\v r'$ of each bond: $A_{\v r\to\v r'}=-A_{\v r'\to\v r}$. Since the lattice is bipartite, all bonds can be oriented consistently from one sublattice towards the other, giving rise to a natural and consistent sign convention.

Scalar fields can also be introduced. They live on the sites of the lattice and, like scalar fields in $\mathbb{R}^d$, their sign is uniquely defined. The lattice gradient of a scalar field and the lattice divergence of a vector field can then be written as
\begin{align}
    (\grad U)_{\v r\to\v r'} &= U_{\v r'}-U_{\v r}\\
    (\div A)_{\v r} &= \sum_{{\v r'}\,\textrm{s.t.} \, \langle \v r \v r'\rangle} A_{\v r\to\v r'}
    \, . 
\end{align}
It is easy to see that the gradient is a well-defined vector field while the divergence is a scalar field. 

The spins in quantum spin ice live on a pyrochlore lattice formed by the bond midpoints of a bipartite diamond lattice (cf.~Fig.~\ref{fig:hexagons}). By employing the sign convention mentioned above, $\sigma^z$ and the $xy$ phase angle of the spins can be turned into a lattice electric field and vector potential, respectively~\cite{Hermele2004PyrochloreMagnet}. A peculiarity of the pyrochlore lattice is that the centres of its hexagonal plaquettes form a dual pyrochlore lattice, whose dual is in turn the original lattice. This allows for defining the lattice curl as the sum of a vector field around a hexagonal plaquette,  traversed according to the right hand rule with respect to the direction of the dual diamond lattice bond corresponding to that plaquette. Upon flipping this direction, the sense of circulation is changed, which in turn flips the sign of all the terms involved. Therefore, the lattice curl is a well-defined vector field of the dual lattice. Furthermore, the diamond sites around each plaquette alternate between the two sublattices, and the lattice curl includes the field variables with alternating signs as compared to the unique sign convention. Ultimately, this leads to the familiar vector calculus identities $\div\curl A=0$ and $\curl\grad U = 0$. 
%
%

\section{Quadratic estimates of the vison cost and interaction}
\label{appendix: interaction}

Following Ref.~\cite{Kwasigroch2017SemiclassicalIce}, we estimate the zero-temperature gap and interaction strength of two visons in a quadratic theory where the energy of the magnetic field on each plaquette, $-g\cos B$, is approximated as $-g+gB^2/2$. Let the number of visons on each site of the dual diamond lattice be $n(\v r)$. We want to find the lowest energy configuration of the magnetic field for which $\div B = 2\pi n(\v r)$.
To do so, we consider a lattice version of Helmholtz's theorem: Any vector field $B(\v r)$ on the dual pyrochlore lattice with periodic boundary conditions can be written as
\begin{equation}
    B=-\grad U+\curl A+C^\mu 
    \, ,
    \label{eq: interaction: Helmholtz}
\end{equation} 
where $U$ is a scalar potential defined on the dual diamond lattice, $A$ is a vector potential defined on the original pyrochlore lattice (both with the same periodic boundary conditions as $B$), and $C^\mu$ is a constant that can vary amongst the four fcc sublattices $\mu=0,1,2,3$ of the dual pyrochlore lattice.
Now, the cross-terms between the three components vanish in $\sum B^2$, that is, the approximate total energy of a field configuration can be written as
\begin{align}
    E-E_0 &\simeq \frac{g}2\sum_{\v r} B^2 =  \frac{g}2\sum_{\v r} \left[(\grad U)^2+(\curl A)^2+(C^\mu)^2\right].
\end{align}
Since $U$ is uniquely determined by $n$, $E$ is minimised if $\curl A$ is identically zero, that is, there are no non-gradient components of $B$. 
%
%

\subsection{Coulomb contribution} 
\label{appendix: interaction: coulomb}
We focus initially on the first two terms in Eq.~\eqref{eq: interaction: Helmholtz} and we shall return to $C^\mu$ later in Sec.~\ref{appendix: interaction: ewald surface}. 
The scalar potential $U$ can be obtained by requiring that
\begin{equation}
    \div B = -\div\grad U = 2\pi n 
    \, .
    \label{eq: interaction: Laplace real space}
\end{equation} 
The divergence and the gradient can be expressed in reciprocal space as 
\begin{align}
    (\grad U)^\mu (\v q) &= 
      -\left[ \v M^\dagger(\v q) \right]^{\mu\alpha} U^\alpha (\v q)\\
    (\div B)^\alpha (\v q) &= 
      \left[ \v M(\v q) \right]^{\alpha\mu} B^\mu (\v q)
      \, ,
\end{align}
where the indices $\alpha=\pm$ and $\mu=0,1,2,3$ indicate the fcc sublattices of the diamond and pyrochlore lattices, respectively, and
\begin{equation}
    \v M(\v q) = \left(\begin{array}{cccc}
        e^{i\v q\cdot\v r_0/2} &e^{i\v q\cdot\v r_1/2} &e^{i\v q\cdot\v r_2/2} &e^{i\v q\cdot\v r_3/2} \\
        -e^{-i\v q\cdot\v r_0/2} &-e^{-i\v q\cdot\v r_1/2} &-e^{-i\v q\cdot\v r_2/2} &-e^{-i\v q\cdot\v r_3/2} 
    \end{array}\right),
\end{equation}
where $\v r_0 = a_0[111]/4$, $\v r_1 = a_0[1\overline{11}]/4$, $\v r_2 = a_0[\overline{1}1\overline{1}]/4$, and $\v r_3=a_0[\overline{11}1]/4$ are the vectors pointing from a ``+'' diamond lattice site to its nearest neighbours. It then follows that~\eqref{eq: interaction: Laplace real space} can be written as (we suppress the arguments $\v q$ from now on)
\begin{align}
    \v M\v M^\dagger \v U &= 2\pi\v n\nonumber\\
    \v U &= 2\pi\big(\v M\v M^\dagger\big)^{-1}\v n \nonumber\\
    \v B = \v M^\dagger\v U &= 2\pi\v M^\dagger \big(\v M\v M^\dagger\big)^{-1}\v n \, , 
    \label{eq:interaction: Laplace reciprocal}
\end{align}
where we introduced the vectors $\v n = [n^+,n^-]$, $\v U = [U^+,U^-]$, and $\v B = [B^0,B^1,B^2,B^3]$ for convenience.
Equation~\eqref{eq:interaction: Laplace reciprocal} gives all $\v q\neq 0$ Fourier components of the ground state magnetic field configuration. [$\v M\v M^\dagger$ is singular at the $\Gamma$ point: this is the origin of the constant term in \eqref{eq: interaction: Helmholtz}.] Now, the total energy of this configuration in the quadratic approximation is
\begin{align}
    E-E_0 &= \frac{g}2\sum_{\v r} B^2(\v r) = \frac{g}{2N}\sum_{\mu,\v q} \big|B^\mu(\v q)\big|^2 = \frac{g}{2N}\sum_{\v q} \v B^\dagger\v B\nonumber\\
    &\simeq \frac{(2\pi)^2g}{2N}\sum_{\v q\neq 0} \v n^\dagger \big(\v M\v M^\dagger\big)^{-1}\v M\v M^\dagger\big(\v M\v M^\dagger\big)^{-1}\v n \nonumber\\
    &\longrightarrow 2\pi^2gV_\mathrm{cell} \int_{\v q\in\mathrm{BZ}} \frac{\ud^3q}{(2\pi)^3}\v n^\dagger \big(\v M\v M^\dagger\big)^{-1}\v n 
    \, , 
    \label{eq: interaction: general energy}
\end{align}
where $N$ is the number of fcc unit cells of volume $V_\mathrm{cell} = a_0^3/4$. 
We also note that 
\begin{align}
    \v M\v M^\dagger &= 4\left(\begin{array}{cc}
        1 & -\gamma \\
        -\gamma^* & 1
    \end{array}\right)\implies\nonumber\\*
    \big(\v M\v M^\dagger\big)^{-1} &= \frac1{4(1-|\gamma|^2)}\left(\begin{array}{cc}
        1 & \gamma \\
        \gamma^* & 1
    \end{array}\right),
\end{align}
where $\gamma=\frac14\sum e^{i\v q\cdot\v r_\mu}=\cos(q_x/4)\cos(q_y/4)\cos(q_z/4)-i\sin(q_x/4)\sin(q_y/4)\sin(q_z/4)$ \cite{Henley2005Power-lawAntiferromagnets}.

To obtain the interaction energy of two visons from the general form (\ref{eq: interaction: general energy}), consider a positive and a negative vison, both on the ``+'' sublattice, a distance $\v R$ away from each other. That is, let $n(0)=+1$, $n(\v R)=-1$, and $n(\v r)=0$ otherwise. In reciprocal space, this yields $n^+(\v q)=1-e^{i\v q\cdot\v R}$ and $n^-(\v q)=0$. Substituting this into (\ref{eq: interaction: general energy}) gives
\begin{align}
    E-E_0 &=\frac{\pi^2gV_\mathrm{cell}}2 \int_{\v q\in\mathrm{BZ}}  \frac{\ud^3q}{(2\pi)^3} \frac{|1-e^{i\v q\cdot\v R}|^2}{1-|\gamma^2|}\nonumber\\
    &= 2\times\frac{ga_0^3\pi^2}{8} \int_{\v q\in\mathrm{BZ}}  \frac{\ud^3q}{(2\pi)^3} \frac1{1-|\gamma|^2}\nonumber\\* &\phantom{=2}-\frac{ga_0^3\pi^2}4\int_{\v q\in\mathrm{BZ}}  \frac{\ud^3q}{(2\pi)^3} \frac{\cos(\v q\cdot\v R)}{1-|\gamma|^2}.
    \label{eq: interaction: two visons}
\end{align}
The two terms of (\ref{eq: interaction: two visons}) can be regarded as the energies of the two isolated visons and their interaction, respectively. The bare vison energy works out to be $\mu\approx 8.848 g$. For large $\v R$, the second term integrates to zero over most of the Brillouin zone due to the rapidly oscillating cosine factor; the only exception is the vicinity of the $\Gamma$ point where $\gamma=1$, and so the integrand diverges. There, $|\gamma|^2\approx 1-\v q^2a_0^2/16$, and so the integral becomes
\begin{align}
    E_\mathrm{int.} &\approx -4ga_0\pi^2 \int \frac{\ud^3q}{(2\pi)^3} \frac{\cos(\v q\cdot\v R)}{\v q^2} = -\frac{ga_0\pi}R
    \, ,
    \label{eq: interaction: Coulomb}
\end{align}
which demonstrates the effective Coulomb interaction between visons at large distances, arising from a purely short-ranged Hamiltonian. 
%
%

\subsection{Ewald surface term}
\label{appendix: interaction: ewald surface}
In the discussion above, we did not consider the $\v q=0$ term of the Fourier transformed energy (\ref{eq: interaction: general energy}) because it is not uniquely determined by the $\div B=2\pi n$ condition. This ambiguity is manifest in the constants $C^\mu$ in the Helmholtz decomposition (\ref{eq: interaction: Helmholtz}).
To derive these constants, we have to note that the origin of visons is a $2\pi$ phase ambiguity in the emergent vector potential $\phi$, the curl of which is the magnetic field $B$. We can fix this ambiguity by giving each $\phi$ a unique value and by introducing an integer-valued field $\mathcal{B}$ that ensures $|B|<\pi$ for each plaquette \cite{Hermele2004PyrochloreMagnet}:
\begin{equation}
    B = \curl\phi + 2\pi\mathcal{B}.
    \label{eq: interaction: Dirac string}
\end{equation}
As $\phi$ is now uniquely defined, $\div\curl\phi = 0$, and so
\begin{equation}
    \div B = 2\pi\div\mathcal{B} \quad \implies \quad q_v = \div\mathcal{B}.
\end{equation}
This means that visons can only be introduced by changing $\mathcal{B}$ on some plaquettes, that is, by phase slips.

It is easy to see that the lattice curl of any field with wave vector $\v q=0$, that is, one that is constant on each pyrochlore sublattice, is uniformly zero. 
As a result, the $\v q=0$ component of the magnetic field~\eqref{eq: interaction: Dirac string} is entirely due to the $2\pi\mathcal{B}$ term, and we now show that it depends only on the position of the visons. 
To do so, consider introducing a pair of visons with (positive to negative) separation $\Delta\v r$. This involves adding $\pm 1$ to $\mathcal{B}$ on plaquettes along some path between the two end points: Heading from the positive to the negative vison, a plaquette encountered going from a ``$+$'' to a ``$-$'' tetrahedron gets a $+1$, while a ``$-$'' to a ``$+$'' one gets a $-1$. Since each of these steps is associated with a vison movement $\pm\v r_\mu$, respectively, we have
\begin{equation}
    \Delta\v r = \sum_\mu \Delta \mathcal{B}^\mu \v r_\mu,
    \label{eq: interaction: change in calli B 1}
\end{equation}
where $\mathcal{B}^\mu$ is the total $\mathcal{B}$ on pyrochlore sublattice $\mu$ (that is, its $\v q=0$ Fourier component). 
Equation~\eqref{eq: interaction: change in calli B 1} gives three equations for the four $\mathcal{B}^\mu$. A fourth one can be obtained by realising that the total $\mathcal{B}$ changes by $+1$ when a vison moves onto the ``$+$'' diamond sublattice and by $-1$ when a vison moves away from it, that is, the total $\Delta \mathcal{B}$ upon inserting a vison pair equals the total change in $q_+$, the net vison charge in the ``$+$'' sublattice:
\begin{equation}
    \sum_\mu \Delta \mathcal{B}^\mu = \Delta q_+.
    \label{eq: interaction: change in calli B 2}
\end{equation}
Since all visons are located at the ends of Dirac strings, one could introduce them from a setup containing no visons by repeating this operation. 
Summing (\ref{eq: interaction: change in calli B 1},\,\ref{eq: interaction: change in calli B 2}) for all of these vison creation events gives
\begin{equation}
    \left[\begin{array}{c}
         -q_+ \\
         4P_x/a_0 \\
         4P_y/a_0 \\ 
         4P_z/a_0 
    \end{array} \right] = -\underbrace{\left[ \begin{array}{rrrr}
        1 & 1 &  1 &  1 \\
        1 & 1 & -1 & -1 \\
        1 &-1 &  1 & -1 \\
        1 &-1 & -1 &  1 
    \end{array}\right]}_{\v N} \left[\begin{array}{c}
         \mathcal{B}^0 \\
         \mathcal{B}^1 \\
         \mathcal{B}^2 \\
         \mathcal{B}^3
    \end{array}\right],
    \label{eq: interaction: change in calli B 3}
\end{equation}
where $\v P = -\sum\Delta\v r$ is the total dipole moment of the vison configuration. From (\ref{eq: interaction: Dirac string},\,\ref{eq: interaction: change in calli B 3}), the contribution of the $\v q=0$ component to the total energy can be written as
\begin{align}
    E_{\v q =0} &= \frac{g}{2N} \sum_\mu |B^\mu(\v q =0)|^2  = \frac{g(2\pi)^2}{2N} \sum_\mu (\mathcal{B}^\mu)^2 \nonumber \\
    &=  \frac{2\pi^2 ga_0}{V} \left[\v P^2 + \left(\frac{q_+ a_0}4\right)^2 \right],
    \label{eq: interaction: k=0 energy}
\end{align}
where $V$ is the volume of the system with periodic boundary conditions and we used that $\v N/2$ is orthonormal.

This result can be connected to the standard surface term in Ewald summation \cite{deLeeuw1980SimulationConstants}, which is of the form
\begin{equation}
    E_\mathrm{surf} = \frac{\kappa\, \v P^2}{2(2\varepsilon+1) V},
    \label{eq: interaction: surface term}
\end{equation}
where $\kappa$ is the Coulomb constant defined by $V_{ij}=\kappa q_iq_j/(4\pi r)$ and $\varepsilon$ is the relative permittivity of the medium surrounding the system. In our case, $\kappa=4\pi^2 ga_0$ and $\varepsilon=0$, for the emergent magnetic field does not even exist outside of quantum spin ice. Substituting this into~\eqref{eq: interaction: surface term} reproduces the $\v P$-dependent part of~\eqref{eq: interaction: k=0 energy}.
%
%

\subsection{Comparison to numerical simulations}

\begin{figure}
    \centering
    \includegraphics{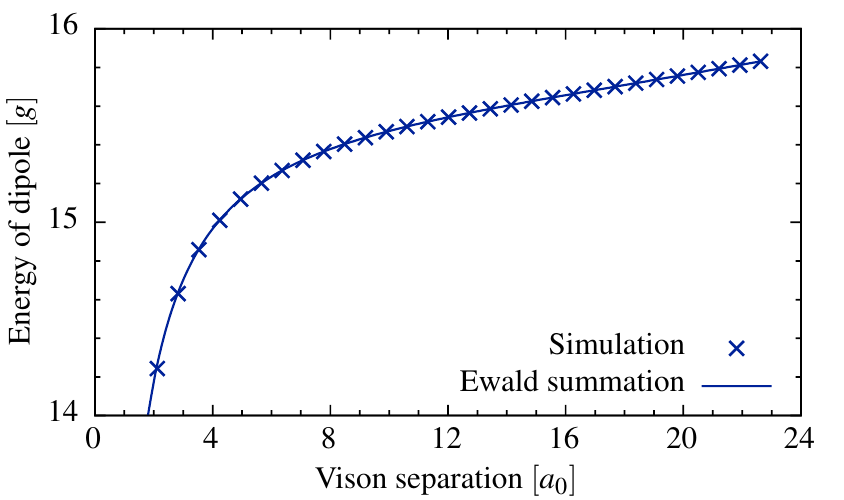}
    \caption{Total energy of a positive and negative vison in the zero-temperature limit, 
    displaced along the $[110]$ lattice direction, in a sample of size $L=32$ unit cells with periodic boundary conditions. 
    Equilibrium energies in the full theory (crosses) are matched perfectly by Ewald summing the effective Coulomb interaction (\ref{eq: interaction: Coulomb},~\ref{eq: interaction: k=0 energy}) on top of the exact chemical potential of two isolated visons (\ref{eq: bare vison: mass, interaction}) (continuous line).}
    \label{fig:ewald}
\end{figure}

In order to confirm the validity of the quadratic theory for interactions, we used the photon cooling protocol described in Sec.~\ref{section: bare vison} to measure the energy of two visons separated by a range of distances in a cubic simulation box of size $32a_0$ in the zero-temperature limit.
The results are plotted in Fig.~\ref{fig:ewald} together with the predictions from Ewald summation of the Coulomb interaction~\eqref{eq: interaction: Coulomb} with  surface term~\eqref{eq: interaction: k=0 energy}, and the bare energy cost of visons taken from~\eqref{eq: bare vison: mass, interaction}. This prediction has no free parameters and agrees excellently with the numerical simulations. 
%
%

\section{Pinch point blurring  due to free and bound visons}
\label{appendix: pinch point}
We derive the contribution of visons to the equal time correlator of the emergent magnetic field, $\langle B(\v q)B(-\v q)\rangle$, for several arrangements of visons that are realised in the semiclassical quantum spin ice model. We assume throughout that there is no interaction between photons and visons, that is, the overall correlator can be written as the sum of independent vison and photon contributions.

Let the vison number on each site of the dual diamond lattice be $n(\v r)$, which can be Fourier transformed into $n^\alpha(\v q)$, as done in Appendix~\ref{appendix: interaction}. 
The contribution of these visons to the magnetic field is given by~\eqref{eq:interaction: Laplace reciprocal}, and therefore
\begin{align}
    B(\v q) = 
    \sum_\mu B^\mu(\v q) = 
    2\pi\v 1\v M^\dagger \big(\v M\v M^\dagger\big)^{-1}\v n(\v q) 
    = \v v \v n(\v q)
    \, ,
\end{align}
where $\v1$ is the row 4-vector all entries of which are 1, and we introduced the row 2-vector $\v v = 2\pi\v 1\v M^\dagger (\v M\v M^\dagger)^{-1}$. 
The latter can be written as 
\begin{equation}
    \v v = \frac{2\pi}{1-|\gamma|^2} \big[\kappa^*-\gamma^*\kappa,\ \  \gamma\kappa^*-\kappa\big]
    \, ,
\end{equation}
where $\kappa = \frac14\sum_\mu e^{i\v q\cdot\v r_\mu/2}$. Therefore, the vison contribution to the correlator of $B(\v q)$ follows from that of $n(\v q)$ as
\begin{equation}
    \langle B(-\v q)B(\v q)\rangle_{\rm vison} = \sum_{\alpha,\beta} \overline{v^\alpha}v^\beta \left\langle n^\alpha(-\v q)n^\beta(\v q)\right\rangle.
    \label{eq: pinchpoint: correlator}
\end{equation}
Below we derive the correlator of vison number, and hence its contribution to that of the magnetic field, for vison arrangements relevant to our model. The most salient feature of the full $\langle B(-\v q)B(\v q)\rangle$ correlator are the pinch points. For convenience, we focus on the behaviour along the $q_z$ axis, where the photon contribution vanishes~\cite{Henley2005Power-lawAntiferromagnets}, and so the $\langle BB\rangle$ correlator measured along it is entirely due to visons and allows for direct comparison with the result in Eq.~\eqref{eq: pinchpoint: correlator}. 
%
%

\subsection{Debye plasma of dissociated visons}
\label{appendix: pinch point: Debye}

Let the visons interact through the reduced Coulomb interaction $\beta V(r)=Kq_iq_j/(4\pi r)$. Assume that the density $\rho$ of visons is small, that is, both their typical separation and the Debye screening length of the resulting plasma is much larger than the lattice spacing. Accordingly, we focus on the vicinity of $\Gamma$ points in reciprocal space. From standard Debye--H\"uckel theory \cite{Levin2002ElectrostaticBiology}, the pair correlation function of visons is 
\begin{align}
    g_{ab}(r) &= e^{-\beta q_aq_b \phi(r)} \approx 1-\beta q_aq_b \phi(r)
    \label{eq: pinchpoint: DH pair corr}\\
    \beta\phi(r) &= \frac{K e^{-r/\xi_\mathrm{D}}}{4\pi r},
\end{align}
where $\xi_\mathrm{D}=(K\rho)^{-1/2}$ is the Debye screening length and $a,b=\pm$ denote the positive and negative vison species. To go from (\ref{eq: pinchpoint: DH pair corr}) to the correlators $\langle n^\alpha n^\beta\rangle$, we note that the long-wavelength theory does not discriminate between the two sublattices. Therefore, the density of visons in each is $\rho/2$ and $\langle n^\alpha n^\beta\rangle$ does not depend on $\alpha$ and $\beta$ in the long wavelength limit. In this approximation,
\begin{align}
    \left\langle n^\alpha(\v R) n^\beta(\v R+\v r)\right\rangle &\propto \left(\frac{\rho}4\right)^2 \sum_{a,b=\pm} q_aq_bg_{ab}(r) =- \frac{\rho^2}4 \frac{Ke^{-r/\xi_\mathrm{D}}}{4\pi r}\nonumber\\
    \left\langle n^\alpha(-\v q)n^\beta(\v q)\right\rangle_2 &=- \frac{K\rho^2}4 \frac1{\v q^2+\xi_\mathrm{D}^{-2}} = -\frac{\rho}4 \frac{\xi_\mathrm{D}^{-2}}{\v q^2+\xi_\mathrm{D}^{-2}}.
    \label{eq: pinchpoint: DH corr 1}
\end{align}

The above derivation captures the correlation between pairs of two visons; however, the correlation of visons with themselves also contributes to the $\langle nn\rangle$ correlator. This contribution is clearly a $\delta$-function in real space, and so only couples each sublattice to itself; after Fourier transforming, we get
\begin{equation}
    \left\langle n^\alpha(-\v q)n^\beta(\v q)\right\rangle_1 = \frac{\rho}2 \delta^{\alpha\beta}.
    \label{eq: pinchpoint: DH corr 2}
\end{equation}

We now calculate the magnetic field correlator $\langle B(-\v q)B(\v q)\rangle $ using (\ref{eq: pinchpoint: correlator}) along the $q_z$ axis near the $\v G = (002)$ pinch point. Since $\v G$ is a $\Gamma$ point of the fcc lattice, both sublattices of the diamond lattice behave the same way as at $\v q=0$, but a relative phase $\v G\cdot\v r_\mu=\pi$ is introduced between them. In practice, this means that the $\langle n^\pm n^\mp\rangle$ correlators are the opposite of what they were near $\v q=0$, while the sign of $\langle n^\pm n^\pm\rangle$ remains unaffected. Equations~(\ref{eq: pinchpoint: correlator},\,\ref{eq: pinchpoint: DH corr 1},\,\ref{eq: pinchpoint: DH corr 2}) now yield
\begin{equation}
    \langle B(-\v q)B(\v q)\rangle_{\rm free} 
    = \frac{(2\pi)^2\rho}{4\sin^2(k/8)} \frac{k^2}{k^2+\xi^{-2}_\mathrm{D}} \approx\frac{64\pi^2}K \frac1{1+\xi^2_\mathrm{D}k^2},
    \label{eq: pinchpoint: DH result}
\end{equation}
where $\v q=(0,0,4\pi/a_0+k)$. That is, the Debye plasma of visons introduces a Lorentzian blurring of the pinch points. The width of this blurring is $\xi^{-1}_\mathrm{D}\propto\sqrt{\rho}$.
%
%

\subsection{Tightly bound dipoles}
\label{appendix: pinch point: dipoles}

Since the dominant vison species at low temperatures is not the isolated vison, but a dipole of nearest neighbour visons, we need to derive $\langle B(-\v q)B(\v q)\rangle$ for a gas of nearest neighbour dipoles. Furthermore, the energy of a second neighbour dipole is significantly smaller (by about $0.48g$) than predicted by the simple Coulomb approximation, see Fig.~\ref{fig:freeze}(b). For these reasons, the Debye plasma approximation will significantly underestimate the population of such dipoles, which must be corrected for explicitly. The only effect of these associated dipoles is correction to the polarisability of the emergent magnetic field. Therefore, their interactions with each other can be neglected, and the only contribution to $\langle nn\rangle$ correlators comes from visons within the same dipole.

For nearest neighbour ($\v r_1-\v r_2 = a_0\langle 111\rangle/4$) dipoles, the real space vison correlators are thus
\begin{equation}
    \langle n(\v r)n(\v r') \rangle \propto \left\{ \begin{array}{ll}
        2\rho & \v r = \v r' \\
        -\rho/4 & \v r'-\v r =a_0\langle 111\rangle/4\\
        0 & \textrm{otherwise,}
    \end{array}\right.
    \label{eq: pinchpoint: 1nn real space}
\end{equation}
where $\rho$ is the density of dipoles; the factor of $1/4$ is due to the four possible orientations of the dipole.
Clearly, the two visons are on the same sublattice in the first line in \eqref{eq: pinchpoint: 1nn real space} and on different ones in the second. 
Therefore, the reciprocal space correlators are
\begin{equation}
    \left\langle n^\alpha(-\v q) n^\beta(\v q)\right\rangle = \rho \left(\begin{array}{cc}
        1 & -\gamma^* \\
        -\gamma & 1
    \end{array}\right)^{\alpha\beta}
    \, ,
\end{equation}
from which the magnetic field correlator follows as
\begin{equation}
    \langle B(-\v q)B(\v q)\rangle_{\rm nn} = \frac{\rho\pi^2}4\frac{2|\kappa|^2-\gamma^*\kappa^2 - \gamma\kappa^{*2}}{1-|\gamma|^2},
    \label{eq: pinchpoint: 1nn result}
\end{equation}
which is constant along the $q_z$ axis. That is, the contribution of nearest neighbour dipoles to the magnetic field correlator is a uniform background that gradually submerges the pinch point, eventually washing it out.

For second neighbour dipoles ($\v r_1-\v r_2 = a_0\langle 110\rangle/2$), the real space vison correlators are
\begin{equation}
    \langle n(\v r)n(\v r') \rangle \propto \left\{ \begin{array}{ll}
        2\rho & \v r = \v r' \\
        -\rho/6 & \v r'-\v r =a_0\langle 110\rangle/2\\
        0 & \textrm{otherwise.}
    \end{array}\right.
\end{equation}
Since the two components of the dipole are on the same sublattice, the reciprocal space correlators are
\begin{align}
    \left\langle n^\alpha(-\v q) n^\beta(\v q)\right\rangle &= \rho\, \bigg(1-\overbrace{\frac13\sum_{i<j} \cos\frac{q_ia_0}2\cos\frac{q_ja_0}2}^\Phi \bigg)\, \delta^{\alpha\beta}\\
    \langle B(-\v q)B(\v q)\rangle_{\rm 2nn} 
    &= 8\rho\pi^2 \frac{(1-\Phi)|\gamma^*\kappa-\kappa^*|^2}{\big(1-|\gamma|^2\big)^2},
    \label{eq: pinchpoint: 2nn result}
\end{align}
which is proportional to $\sin^2(q_za_0/8)$ along the $q_z$ axis. This explains the small but significant cosine modulation of the lifting of the pinch point.

\begin{figure}
    \centering
    \includegraphics{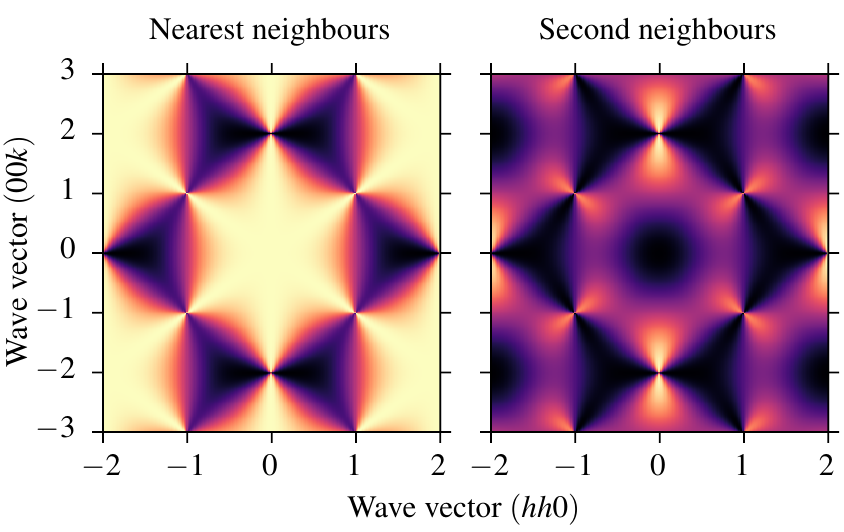}
    \caption{Static correlation function $\langle B(-\v q)B(\v q)\rangle$ due to uniformly distributed bound pairs of visons at nearest-neighbour (left) and next-nearest-neighbour (right) distance. In the first case, the pinch point pattern is complementary to that due to photons [see Fig.~\ref{fig:pinch_points}(b)]. For next-nearest neighbours, the correlator is largest near the pinch points. This behaviour is reminiscent of the Debye plasma forming at large vison separation which gives rise to Lorentzian peaked correlators in the immediate vicinity of pinch points. 
    }
    \label{fig:nnn_correlators}
\end{figure}

Besides the $q_z$ axis, we considered the behaviour of the $\langle BB\rangle$ correlators around the pinch points, especially the one at $\v G=(002)$. Vison number correlators are analytic in all cases; however, the ``response function'' $\v v$ gives rise to pinch points perpendicular to those due to photons (see Fig.~\ref{fig:pinch_points}). The Debye plasma contribution is peaked at the pinch point. Its maximum matches the intensity of the photon contribution there, resulting in Lorentzian pinch point blurring \cite{Henley2005Power-lawAntiferromagnets,Sen2013CoulombDisorder}. On the other hand, closely associated dipoles result in contributions proportional to the density throughout reciprocal space, which reduce the overall contrast of the pinch points. For completeness, the $\langle BB\rangle$ correlators produced by first and second neighbour dipoles,~\eqref{eq: pinchpoint: 1nn result} and~\eqref{eq: pinchpoint: 2nn result}, are plotted on the $(hhk)$ plane in Fig.~\ref{fig:nnn_correlators}. 
%
%

\section{Estimates of the vison gap from thermal statistics}
\label{appendix: gap estimate}
We discuss two methods to estimate the chemical potential of visons in a thermodynamic set-up. We also derive an expression for both of them in terms of statistics of the vison number and the energy at a single temperature, in a similar fashion to the fluctuation--dissipation theorem.
We treat our model as a classical thermodynamic system with microstates of well-defined energy $E_\alpha$ and vison number $N_\alpha$, weighted according to the partition function
\begin{equation}
    Z = \sum_\alpha e^{-\beta E_\alpha+\zeta N_\alpha},
    \label{eq: gap estimate: partition function}
\end{equation}
where $\zeta$ is a fictitious chemical potential introduced to keep track of vison number; in the physical partition function, $\zeta=0$.

The first approach estimates the energy cost of a single vison directly, which can formally be written as $\mu_E = \ud E(N)/\ud N$, where $E(N)$ is the mean energy of the system constrained to $N$ visons. In the thermodynamic limit, this derivative is equivalent to the ratio of variations in $\langle E\rangle$ and $\langle N\rangle$ due to a changing chemical potential:
\begin{equation}
    \mu_E
    \xrightarrow{N\to\infty} \frac{\partial E(\zeta)/\partial\zeta}{\partial N(\zeta)/\partial \zeta}.
    \label{eq: gap estimate: equivalence}
\end{equation}
Each of the two derivatives in (\ref{eq: gap estimate: equivalence}) can be expressed in terms of derivatives of $Z$, which can in turn be rewritten in terms of statistics of energies:
\begin{align}
    \left.\frac{\partial E}{\partial\zeta}\right|_\beta &= -\frac{\partial^2\log Z}{\partial \beta\partial\zeta} = \frac1{Z^2} \frac{\partial Z}{\partial \beta}\frac{\partial Z}{\partial\zeta} - \frac1Z \frac{\partial^2Z}{\partial\beta\partial\zeta} \nonumber\\*
    &= \langle NE\rangle - \langle N\rangle\langle E\rangle = \cov(N,E)\,, 
    \\
    \left.\frac{\partial N}{\partial\zeta}\right|_\beta &= \left.\frac{\partial^2\log Z}{\partial\zeta^2}\right|_\beta = \frac1Z\left.\frac{\partial^2 Z}{\partial\zeta^2}\right|_\beta - \left(\frac1Z\left.\frac{\partial Z}{\partial\zeta}\right|_\beta\right)^2 \nonumber\\*
    &= \langle N^2\rangle - \langle N\rangle^2 = \var N\,,
    \label{eq: gap estimate: muE steps}
\end{align}
and therefore 
\begin{align}
    \mu_E &= \frac{\cov(E,N)}{\var N}\,.
    \label{eq: gap estimate: muE result}
\end{align}

Another estimate of the excitation gap is the local slope of the Arrhenius plot $\log N$ vs.\ $1/T$, at least at low temperatures where $N$ has not saturated. However, this is not necessarily the gap of a single vison, but  of whatever (possibly multi-vison) excitations are created thermally in the system. This slope is given by
\begin{align}
    \mu_\mathrm{Arrh.} &= -\frac{\ud \log N}{\ud\beta} = -\frac1N \frac{\ud N}{\ud\beta} = -\frac1N \frac{\partial^2\log Z}{\partial \beta\partial\zeta}\nonumber\\*
    &= \frac{\cov (E,N)}{N}.
\end{align}

Finally, it is instructive to consider $\mu_\mathrm{Arrh.}/\mu_E = \var N/N$. As discussed before, if the dominant thermal (collective) excitation of the system consists of $m$ visons, $\mu_\mathrm{Arrh.}$ is expected to be $m\mu_E$ at low temperatures, and so $\var N/N\approx m$. We can obtain this last result directly by considering that the number $\tilde{N}$ of collective excitations obeys a Poisson distribution at low temperatures and so $\var \tilde{N}=\tilde{N}$. The result then follows from $N=m\tilde{N}$.
%
%

\section{\label{appendix: toy partition function}
Semiquantitative model of the partition function}

\begin{figure}
    \centering
    \includegraphics{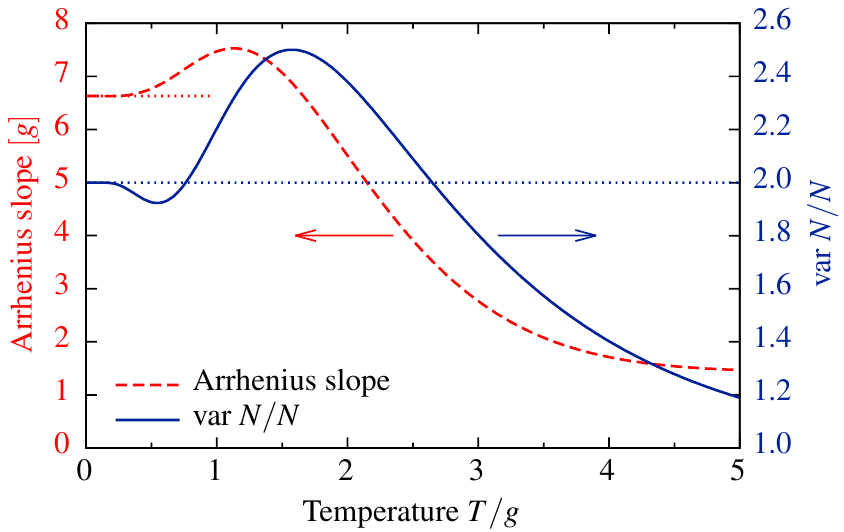}
    \caption{Effective Arrhenius gap $\mu_\mathrm{Arrh.}=-\ud\log N/\ud\beta$ (red dashed line) and $\var N/N$ (blue solid line) of visons in the semiquantitative model discussed in Appendix~\ref{appendix: toy partition function}. The zero-temperature energy cost of bare visons and nearest-neighbour vison pairs, $E_1$ and $E_2$, are given by the low-temperature simulations in Sec.~\ref{section: bare vison}. 
    The model yields the expected zero-temperature limits of both quantities, as well as a qualitative temperature dependence similar to that seen in Fig.~\ref{fig:gap}, for a wide range of the phenomenological parameters $E_\mathrm{int}(T=0)$ and $m$. They were chosen by hand to be $5g$ and 20, respectively, in order to achieve a good numerical agreement.}
    \label{fig:toy_part}
\end{figure}

The behaviour of vison number found in the simulations can be explained by the following three observations:
\begin{enumerate}
    \item The energy cost of a nearest-neighbour dipole of visons is smaller than that of a single isolated vison.
    \item An isolated vison polarises its surroundings, reducing the energy cost of nearby, aligned vison dipoles even further.
    \item The energy cost of visons reduces as temperature increases because highly excited photon modes ``wash them out.''
\end{enumerate}
To demonstrate this, we construct a simplistic model of the partition function~\eqref{eq: gap estimate: partition function}, which is semiquantitative at low temperatures and captures the salient features at large $T$. If we ignore interactions, other than the ones within two visons in a pair, the partition function factorises by diamond lattice sites: $Z=Z_1^\mathcal{V}$, where $\mathcal{V}$ is the number of such sites. Now, each site can host an isolated vison of either charge, or one half of a nearest-neighbour dipole. Assuming their energy cost is $E_1$ and $E_2$, respectively, we can write down a first approximation to $Z_1$ as
\begin{equation}
    Z_1\approx 1+2e^{-\beta E_1+\zeta}+4e^{-\beta E_2+2\zeta},
\end{equation}
where $E_2<E_1$ in line with observation 1. 

Observation 2 states that the energy cost $E_2$ of a dipole near an isolated vison with the right alignment is reduced to $E_2-E_\mathrm{int}$ while that of a dipole on the same bond but with opposite orientation is increased to $E_2+E_\mathrm{int}$. This changes the partition function of the bond by 
\begin{align}
    \Delta Z_1 &\approx e^{2\zeta}\left(e^{-\beta (E_2+E_\mathrm{int})}+e^{-\beta (E_2-E_\mathrm{int})}-2e^{-\beta E_2}\right) \nonumber\\*
    &= 2e^{-\beta E_2+2\zeta}\big[\cosh(\beta E_\mathrm{int})-1\big].
    \label{eq: toy partition function: delta Z} 
\end{align}
In a crude approximation, we assume that each isolated vison introduces a fixed $E_\mathrm{int}$ to $m$ nearby bonds and that the resulting $\Delta Z_1$ can be factored into the partition function of the isolated vison. This gives
\begin{equation}
    Z_1\approx 1+4e^{-\beta E_2+2\zeta}+2e^{-\beta E_1+\zeta} (1+\Delta Z_1)^m.
    \label{eq: toy partition function: Z}
\end{equation}

Observation 3 concerns the strong interactions between a highly excited photon bath and the visons, a full treatment of which is a tall order. To estimate its effect on the vison thermodynamics, we propose a ``Hartree--Fock approximation'' where the effect of visons on the photon cloud is neglected, and we assume that the quadratic theory governing the photon modes at low temperatures~\cite{Kwasigroch2017SemiclassicalIce} holds at arbitrary $T$. In this approximation, the energy associated with the gradient component $B_\mathrm{grad}$ of the magnetic field is 
\begin{align}
    E(B_\mathrm{grad},T) &= -g \langle\cos(B_\mathrm{grad}+B_\mathrm{curl})\rangle 
    = -g\cos (B_\mathrm{grad}) e^{-T/(4g)}\,, 
    \label{eq: toy partition function: renormalisation}
\end{align}
since $B_\mathrm{curl}$ on each site has a Gaussian distribution of variance $T/(2g)$. That is, all energy scales associated with $B_\mathrm{grad}$, and hence with the visons, are exponentially suppressed at high temperatures. 

The partition function (\ref{eq: toy partition function: delta Z}--\ref{eq: toy partition function: renormalisation}) can now be used to derive thermodynamic quantities. We plotted $\var N/N$ and the slope of the Arrhenius curve $\ud(\log N)/\ud\beta$ in Fig.~\ref{fig:toy_part}; the results are in good qualitative agreement with the numerical simulation, cf.~Fig.~\ref{fig:gap}. Saturation occurs at a higher temperature than in the full large-$S$ treatment, albeit well below the zero-temperature energy cost of visons. More accurate estimates would likely follow from taking the emergent electric field and photon--photon interactions into account, but this is beyond the scope of the present paper.

\vfill
%
%

\bibliography{references,crossref}

\end{document}